\documentclass[12 pt]{article}

\usepackage{graphicx}
\usepackage{amsmath}
\usepackage{xcolor}
\usepackage{amssymb}
\usepackage{wrapfig}
\usepackage[margin=0.9in]{geometry}
\usepackage{makecell}
\usepackage[normalem]{ulem}

\usepackage{caption}
\usepackage{setspace}
\usepackage[numbers]{natbib}

\newcommand{\nhat}{\mathbf{\hat{n}}}
\newcommand{\Nhat}{\mathbf{\hat{N}}}

\newcommand{\nhatZ}{\mathbf{\hat{n}_{0}}}

\newcommand{\rvec}{\mathbf{r}}
\newcommand{\Rvec}{\mathbf{R}}
\newcommand{\ohat}{\mathbf{\hat{o}}}

\newcommand{\dn}{\delta \mathbf{n}}

\newcommand{\dny}{\delta n_{y}}

\newcommand{\xhat}{\mathbf{\hat{x}}}
\newcommand{\yhat}{\mathbf{\hat{y}}}
\newcommand{\zhat}{\mathbf{\hat{z}}}

\newcommand{\um}{\,\mu\text{m}}
\newcommand{\bfr}{\mathbf{r}}
\newcommand{\nn}{\text{nn}}
\newcommand{\dd}{\text{d}}

\newcommand{\nChr}{n_{\text{chr}}}

\usepackage{authblk}
\author[1]{Colm P.~Kelleher}
\author[2]{Yash Rana}
\author[1,2,3]{Daniel J.~Needleman}
\affil[1]{Department of Molecular and Cellular Biology, Harvard University, Cambridge, MA 02138}
\affil[2]{John A.~Paulson School of Engineering and Applied Sciences, Harvard
  University, Cambridge, MA 02138}
\affil[3]{Center for Computational Biology, Flatiron Institute, New York, NY 10010}

\begin{document}

\title{Long-Range Repulsion Between Chromosomes in Mammalian Oocyte Spindles}
\date{\today}

\maketitle

\newcounter{SInumber}
\setcounter{SInumber}{1} 

\newcommand{\initializeSIsec}[1]{%
    \edef#1{\theSInumber}%
    \stepcounter{SInumber}%
  }

\newcommand\headings{0}

\textbf{During eukaryotic cell division, a microtubule-based structure called the spindle
exerts forces on chromosomes, thereby organizing and segregating
them~\cite{AlbertsBook2016}. Extensive work demonstrates that the forces acting parallel
to the spindle axis, including those responsible for separating sister chromatids, are
generated by microtubule polymerization and depolymerization, and molecular-motors
~\cite{Maiato2017, Anjur-Dietrich2021, Pavin2016, Nazockdast2020}. In contrast, little is
known about the forces acting perpendicular to the spindle axis, which determine the
configuration of chromosomes at the metaphase plate, and thus impact nuclear
localization and rates of segregation errors~\cite{Gerlich2003, Klaasen2022}. Here,
we use quantitative live-cell microscopy to show that metaphase chromosomes are spatially
anti-correlated in mouse oocyte spindles, indicating the existence of hitherto unknown
long-range forces acting perpendicular to the spindle axis. We explain this observation by
first demonstrating that the spindle's microtubule network behaves as a nematic liquid
crystal, and then arguing that deformation of the nematic field around embedded
chromosomes causes long-range repulsion between them. Our work highlights the surprising
relevance of materials physics in understanding the structure, dynamics, and mechanics of
cellular structures, and presents a novel and potentially generic mode of chromosome
organization in large spindles.}

\ifnum \headings > 0
\section{Steady-State Microtubule Orientation in Oocyte Spindles
  Follows Circle Arcs}
\fi

Chromosome segregation is a physical and mechanical process, requiring precisely
coordinated motion of micron-sized objects (chromosomes) through distances of tens of
microns (the size of a typical metazoan cell)~\cite{AlbertsBook2016,
Anjur-Dietrich2021}. The forces causing this motion are generated primarily by the
spindle, a cellular structure comprising a network of microtubules -- long, rigid polymers
of the protein tubulin -- in association with hundreds of additional proteins that
modulate microtubule nucleation, polymerization/depolymerization, and
interactions~\cite{Eggert2006, Mogessie2018}. However, despite our extensive knowledge of
the spindle's molecular constituents, we lack a general framework for understanding how it
self-organizes to generate cellular-scale forces. This places a fundamental limit on our
ability to predict when spindle dysfunction leads to errors in chromosomes
segregation, and thus how it might contribute to diseases such as cancer and
infertility~\cite{Silkworth2012, Holubcova2015, Potapova2017}.

\begin{figure}[t!]
  \centering
  \includegraphics[width=\textwidth]{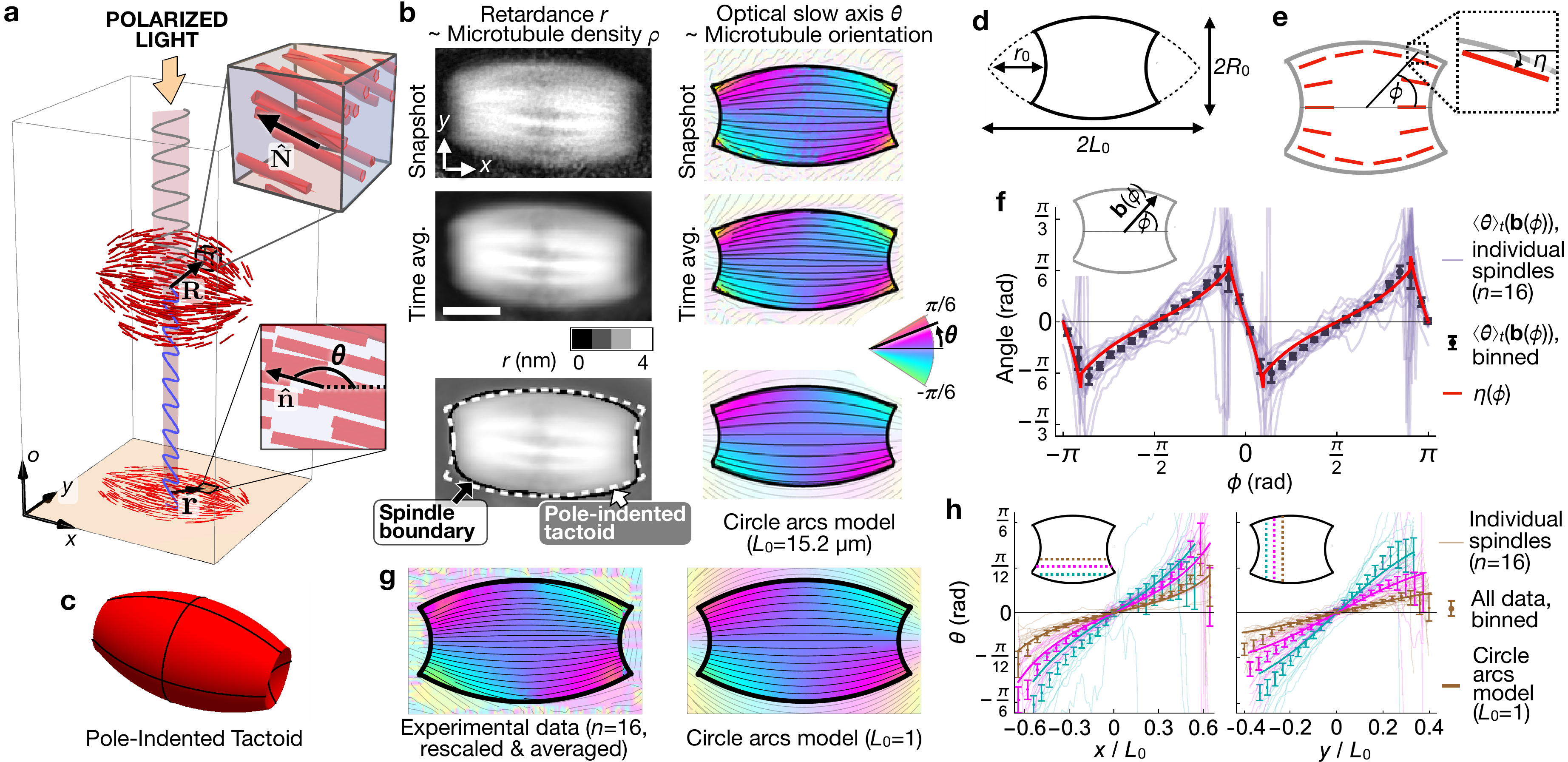}
\caption{\label{fig:polscope} In living mouse oocyte spindles, steady-state microtubule
orientation is fully determined by surface geometry.  (a) For a spindle oriented with its
long axis $\xhat$ perpendicular to the optical axis $\ohat$, the retardance $r$ is related
to the cross-sectional density $\rho(\Rvec)$ of microtubules integrated over $\ohat$. The
coarse-grained 3D microtubule orientation is described by a field of unit vectors
$\Nhat(\Rvec)$; at each pixel of a 2D LC-PolScope image, the optical slow axis angle
$\theta(\rvec)$ specifies $\nhat(\rvec)$, the normalized projection of $\Nhat(\Rvec)$ over
$\ohat$. (b) \textit{Upper and Middle Rows:} Snapshots and time averages (500 frames, 40
mins) of $r$ and $\theta$ for an MII spindle with $\xhat \perp \ohat$. Scale bar $5
\um$. \textit{Bottom Left:} The spindle boundary (black solid curve) is fitted to a pole-indented tactoid
(white dashed curve) with $L_{0} = 15.2 \um$. \textit{Bottom Right:} Using the same value
of $L_{0}$, the circle arcs model generates an angle field that closely matches the
experimentally observed microtubule orientation. (c) 3D view of a pole-indented tactoid,
representing the average fit parameters of MII spindles (Methods 4). (d) Generatrix of a
pole-indented tactoid, with virtual poles at $(\pm L_{0},0)$. (e) On the
convex [concave] part of the spindle boundary, $\eta(\phi)$ is defined as the angle
tangent [perpendicular] to the best-fit pole-indented tactoid, where $\phi$ is the polar
angle in the $xy$-plane. (f) \textit{Inset:} In polar coordinates, the spindle boundary is
parameterized by the curve $\mathbf{b}(\phi)$. \textit{Main:} Without additional fit
parameters, the measured microtubule orientation at the spindle boundary $\langle \theta
\rangle_{t} (\mathbf{b}(\phi))$ (thin purple curves \& black points with error bars)
closely follows the predicted form $\eta(\phi)$ (red solid curve). (g) Rescaling of
experimental data by pole spacing, $\rvec \rightarrow \rvec/L_{0}$, allows averaging over
spindles (Methods 5). The circle arcs model predicts that rescaled data is well-described by a master
field generated by the tangent angles to family of arcs that intersect at $(\pm 1,0)$,
outside the region shown in these images. (h) Horizontal and vertical line profiles of
microtubule orientation. Faint curves show the line profiles of individual spindles, which
are binned to generate the points with error bars. The predictions of the master circle
arcs model ($L_{0}=1$), which contains no fit parameters, are shown as bold curves. Error bars
in (f) and (h) indicate SE.}
\end{figure}

Metaphase II (MII) spindles in mammalian oocytes provide an ideal model system in which to
study spindle-self organization \textit{in vivo}, since their large sizes and long
steady-state lifetimes facilitate detailed microscopy measurements (Supplemental
Information S.I.~\theSInumber\stepcounter{SInumber},~\cite{Liu2000, Holubcova2015,
Mihajlovic2018}).  To characterize the structure and dynamics of the microtubule network
in these spindles, we acquired LC-PolScope movies
(Fig.~\ref{fig:polscope}(a); Methods 1 \& 2). The LC-PolScope is a label-free quantitative
polarization microscope that simultaneously measures the optical retardance $r(\rvec, t)$
and optical slow axis $\theta(\rvec, t)$ at a given time $t$ and each position
$\mathbf{r}$ in a two-dimensional (2D) image~\cite{Liu2000}. These measurements provide
quantitive information regarding the coarse-grained microtubule cross-sectional density
$\rho(\Rvec,t)$ and the nematic director $\Nhat(\Rvec,t)$, both defined at every point
$\Rvec$ in 3D space as well as in time~\cite{Conway2022}: if the spindle long axis $\xhat$
is perpendicular to the optical axis $\ohat$,
\begin{equation}
r(\rvec, t) \approx A_{0}\int_{T}\rho(\Rvec,t) \, \dd o;  \qquad
\nhat(\rvec, t) \equiv (\cos{\theta(\rvec, t) }, \sin{\theta(\rvec, t)}) \approx
\frac{\int_{T} \Nhat(\Rvec, t) \dd o}{||\int_{T}\Nhat(\Rvec,t) \dd o||},
\label{eq:PolQuantDef}
  \end{equation}
  where the integrals are taken over the optical axis $\ohat$, the constant $A_{0} \approx
7.5$ nm$^{2} $ characterizes the retardance contribution of a single microtubule, $T$ is
the sample thickness along the optical axis, and the 2D vector $\nhat(\rvec,t)$ is the
normalized projection of $\Nhat(\Rvec, t)$ into the LC-PolScope image plane
(Fig.~\ref{fig:polscope}(a);
S.I. ~\theSInumber\stepcounter{SInumber} \& \theSInumber\stepcounter{SInumber};
~\cite{Sato1975,Oldenbourg1998, Oldenbourg2013}).

We first used LC-PolScope movies to determine the relationship between the surface
geometry of MII oocyte spindles and microtubule orientation in the spindle interior. To
characterize surface geometry, we identified spindle boundaries from time-averaged
retardance images, $\langle r \rangle_{t}$, calculated in a spindle-referenced coordinate
system where $\zhat = \ohat$ and $\yhat = \zhat \times \xhat$
(Fig.~\ref{fig:polscope}(b,~\textit{left column}), Methods 2-4). We find that those
portions of the spindle boundary that are furthest from the central axis are well-fitted
by a pair of circle arcs that intersect at ``virtual poles'' at $(\pm L_{0},0)$, outside
the spindle boundary. In 3D, the corresponding portion of the spindle surface is convex,
and approximates a tactoid, the shape generated by rotating a circle arc about its
chord. Since the convex surfaces of MII spindles do not extend all the way to the poles,
we model them as ``polar-indented tactoids,'' truncated tactoids with concave spherical
caps of radius $r_{0}$ replacing the poles (Fig.~\ref{fig:polscope}(c \& d);
S.I.~\theSInumber\stepcounter{SInumber}). In materials physics, tactoid and tactoid-like
droplets are a characteristic feature of nematic liquid crystals, and have been observed
under a wide variety of conditions in both experiments and
simulations~\cite{DeBenedictis2016, Wang2018, Kuhnhold2022}. We next examined how
microtubule orientation at the spindle boundary depends on surface geometry. We found
close agreement between the observed microtubule orientation and a purely geometrical
model in which microtubules lie tangent to the spindle's convex (tactoid) surface, and
perpendicular to its concave (polar cap) surfaces (Fig.~\ref{fig:polscope}(e \& f)). Thus,
at the spindle surface, microtubules obey a ``strong anchoring'' boundary
condition~\cite{deGennes1993}. In tactoid-shaped nematic droplets with strong anchoring,
the director is predicted to lie tangent to the unique family of circle arcs that
intersect at $(\pm L_{0},0,0)$~\cite{Williams1986, Prinsen2003}. To test whether this is
the case in MII spindles, we compared the time-averaged slow axis images $\langle \theta
\rangle_{t}$ to the predictions of the circle arcs model (Methods 5). For individual
spindles, we find good agreement between the observed pattern of microtubule orientation
and that predicted by the model (Fig.~\ref{fig:polscope}(b,~\textit{bottom right});
S.I.~\theSInumber\stepcounter{SInumber}). To further probe the validity of the circle arcs
model, we investigated whether steady-state orientation fields from different spindles
collapse onto a master field when they are properly rescaled, as that model predicts they
should. Using previously measured values of $L_{0}$ (and no additional fit parameters), we
rescaled the orientation field of all spindles, $\langle \theta \rangle_{t}(\rvec)
\rightarrow \langle \theta \rangle_{t} (\rvec/L_{0})$, and observed excellent data
collapse (Fig.~\ref{fig:polscope}(g \& h)). Taken together, these results provide strong
evidence for a model of spindle self-organization in which microtubule orientation is
determined by a well-defined anchoring condition on the spindle boundary, together with a
tendency for microtubules in the spinde interior to locally align with each other, i.e.~nematic elasticity.

\begin{figure}[t!]  \centering
\includegraphics[width=\textwidth]{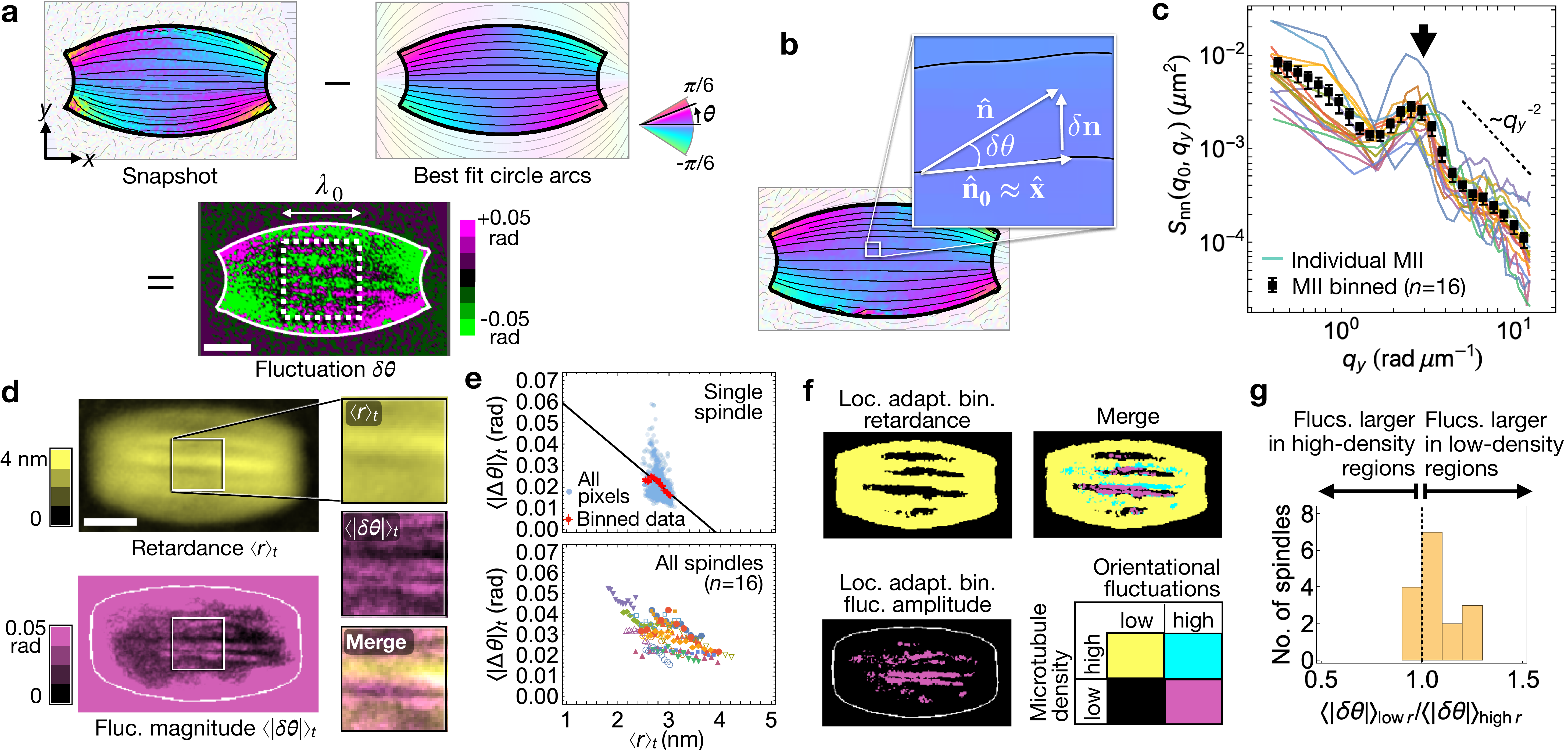}
\caption{\label{fig:flucs} Orientational fluctuations in mouse oocyte spindles are highest
in regions of low microtubule density. (a) Fluctuations are calculated by subtracting
steady-state values of the microtubule orientation field, given by the circle-arcs model,
from instantaneous values. Correlation functions are calculated in the white dashed
box (side length $\lambda_{0}$) in the middle of the spindle,
where $\nhatZ \approx \xhat$.  (b) Near the middle of the spindle, $\dn \cdot \nhatZ
\approx \dn \cdot \xhat = 0 $ and, to lowest order in $\delta \theta$, $\dn = \delta
\theta\, \yhat$.  (c) Spatial correlation function $s_{\nn}(q_{0}, q_{y})$ for 16
spindles, where $q_{0} = 2\pi/\lambda_{0}$ and $q_{y}$ is the wave-vector component perpendicular to the spindle
axis. Correlation functions for individual spindles are plotted in color; binned data
points are obtained by averaging; error bars indicate SE. At the longest and shortest
wavelengths (smallest and largest $q_{y}$), $s_{\nn}$ decays as $q_{y}^{-2}$; around $
q_{y}^{*} = (2.9 \pm 0.1) $ rad $\um^{-1}$ (black arrow; uncertainty estimate in S.I.~6),
a peak appears. (d) Time-averaged retardance $\langle r \rangle_{t}$ (\textit{top},
yellow) and orientational fluctuations $\langle |\delta \theta |\rangle_{t}$
(\textit{bottom}, purple). Scale bar 5 $\um$; spindle boundary shown as thin black curve
in $\langle |\delta \theta |\rangle_{t}$ image; inset shows a $5 \um \times 5 \um$ box at
the center of the spindle. (e) \textit{Top:} Near the center of the spindle (white box in
(d)), orientational fluctuations are negatively correlated with retardance. Black line
shows best fit to binned data, with slope $(-0.017 \pm 0.007)$ rad nm$^{-1}$ (95\% CI).
\textit{Bottom:} Orientational fluctuations are negatively correlated with retardance in
all but one spindle; the average slope is $(-0.010 \pm 0.002)$ rad nm$^{-1}$ (mean $\pm$
SE). (f) Applying a local binarization filter (radius $2 \um$) to $\langle
r \rangle_{t}$ and $\langle |\delta \theta |\rangle_{t}$ reveals elongated regions of low
microtubule density, whose long axes are roughly parallel to the spindle axis;
orientational fluctuations are larger in these low-density regions. All images are masked
to only show features inside the spindle boundary (thin white curve in bottom left
image). (g) Histogram of the ratio of the magnitude of orientational fluctuations in
high-retardance regions and low-retardance regions, as identified by local 
binarization. Orientational fluctuations are on average larger in
lower-density regions; the average value of $\langle |\delta \theta|\rangle_{\text{low }
r} /\langle | \delta \theta | \rangle_{\text{high } r}$ over all spindles is $1.07 \pm
0.02$ (mean $\pm$ SE).
}
\end{figure}

\ifnum \headings > 0
\section{Orientational Fluctuations Are Larger in Regions of Low
  Microtubule Density}
\fi
Since the steady-state orientation of microtubules in MII spindles is well-described by
nematic liquid crystal physics, we next investigated if such a model can also describe the
fluctuations in microtubule orientation around that steady-state. In synthetic materials,
spatiotemporal correlations of fluctuations have long been used to characterize material
properties~\cite{Orsay1969, deGennes1993}. More recently, fluctuation analysis was used to
show that the microtubule network of {\it in vitro} reconstituted \textit{Xenopus} egg
extract spindles behaves as an active nematic material~\cite{Brugues2014}. To determine if
a similar approach can be applied to spindles in living oocytes, we first subtracted the
best-fit 2D director field, $\mathbf{\hat{n}_{0}}(\mathbf{r})$, from the instantaneous
field $\nhat(\bfr,t)$ to calculate the fluctuations $\dn(\bfr, t) = \nhat(\bfr,t) -
\mathbf{\hat{n}_{0}}(\rvec)$ in a box of side length $\lambda_{0} = 8 \um$ placed at the
center of the spindle (Fig.~\ref{fig:flucs}(a)). Fluctuations take a particularly simple
form in this region since, to lowest order in $\delta \theta$, $\dn \approx \dny \yhat$
and $\dny = \delta (\sin{\theta}) \approx \delta \theta$ (Fig.~\ref{fig:flucs}(b)).  To
quantify the fluctuation pattern, we plotted the equal time correlation function,
$s_{\nn}(q_{0}, q_{y})$, as a function of the wave-vector component $q_{y}$ perpendicular
to the spindle axis, with the parallel component fixed at the lowest available mode $q_{0}
= 2\pi/\lambda_{0}$ (Methods 6). For the lowest and highest values of $q_{y}$,
$s_{\nn}(q_{0}, q_{y})$ displays behavior consistent with the inverse square power law
predicted by active nematic theory and observed in previous experiments on reconstituted
\textit{Xenopus} spindles (Fig.~\ref{fig:flucs}(c),
S.I.~\theSInumber\stepcounter{SInumber}). At intermediate value of $q_{y}$, however,
$s_{\nn}(q_{0}, q_{y})$ displays a prominent feature neither predicted by theory nor
observed in \textit{Xenopus} extract: a peak centered at $q_{y}^{*} = (2.9 \pm 0.1)$ rad
$\um^{-1}$, corresponding to a real-space wavelength $\lambda^{*} = 2\pi/q_{y}^{*} = (2.2
\pm 0.1) \um$.

To elucidate the origins of the anomalous behavior of $s_{\nn}(q_{0}, q_{y})$, we next
explored the relationship between orientational fluctuations and microtubule density. We
observed that the time-averaged orientational fluctuation magnitude $\langle |\delta
\theta| \rangle_{t}$ was negatively correlated with the time-averaged retardance $\langle
r \rangle_{t}$ in 15 out of 16 spindles (Fig.~\ref{fig:flucs}(d \& e);
S.I.~\theSInumber\stepcounter{SInumber}). To investigate the basis of this negative
correlation, we used a local binarization filter (Methods 7), which
revealed that spindles contain elongated regions with low microtubule density
(Fig.~\ref{fig:flucs}(f)) in which orientational fluctuations are larger
(Fig.~\ref{fig:flucs}(g)).

\begin{figure}[t!]
  \centering
  \includegraphics[width=17 cm]{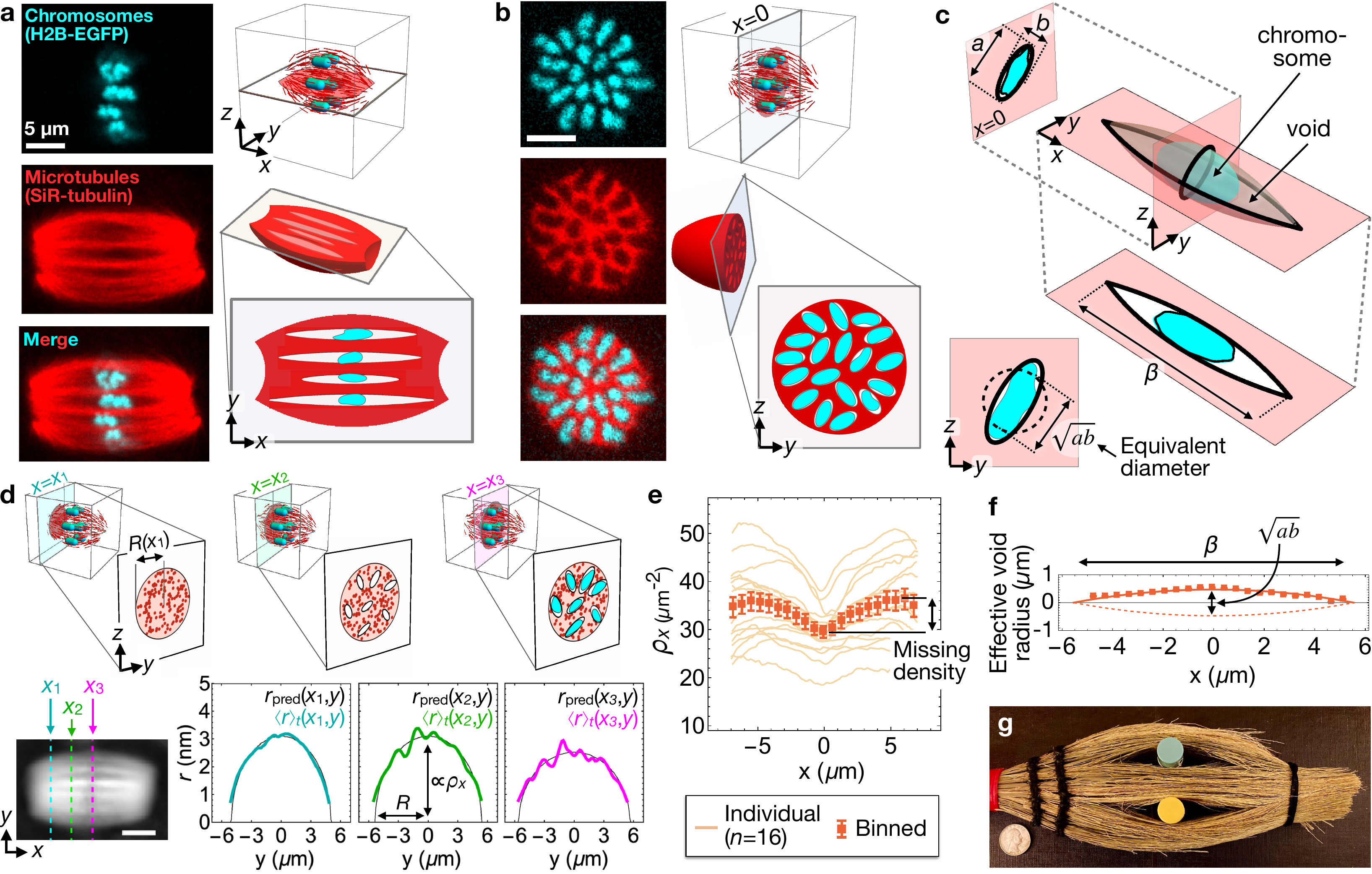}
  \caption{\label{fig:voids} Chromosomes create long, tactoid-like voids in the
microtubule network of mouse oocyte spindles. (a) Confocal micrographs of living oocytes,
taken with the spindle long axis in the confocal plane. Microtubules are labeled with
SiR-tubulin, chromosomes with H2B-EGFP. (b) Micrographs of the metaphase plate ($x=0$
plane). In this plane, chromosome positions (\textit{top image}, cyan) correspond to
regions of low microtubule density, i.e.~voids (\textit{middle image}, black gaps in red
spindle). (c) The typical void has a tactoid-like 3D shape. Along $\xhat$, the void length
$\beta$ is much greater than the chromosome size. In the metaphase plate, the void
cross-section closely matches the chromosome cross-section and may be approximated as an
ellipse with major and minor axes $a$ and $b$. \textit{Bottom Left Inset:} The true
tactoid of equal volume to the void shown has a circular cross section with waist diameter
$\sqrt{ab}$. (d) \textit{Top Row:} For spindles with their long axes perpendicular to the
optical axis, the time-averaged LC-PolScope retardance profile $ r (x_{i},y)$ at a given
position $x_{i}$ along the long axis can be fitted to the formula
$r_{\text{pred}}(x_{i},y) = 2 A_{0} \rho_{x}(x_{i}) (R(x_{i})^{2}-y^{2})^{1/2}$, where
$\rho_{x}(x_{i}) $ is the microtubule cross-sectional density and $R(x_{i})$ the spindle
radius in the plane $x=x_{i}$ (S.I.~3).  Assuming microtubule density is constant outside
the voids, larger values of $\rho_{x}(x_{i})$ correspond to less area taken up by
voids. \textit{Bottom Row:} Fits to $\langle r \rangle_{t}(x_{i}, y)$ at three different
positions $x_{1}$, $x_{2}$, $x_{3}$, along the spindle axis. (e) Best-fit values of
microtubule cross-section density $\rho_{x}$ as a function of position $x$ along the
spindle long axis.  Faint solid curves represent data from individual spindles; points
with error bars represent the average over all spindles. (f) By assuming that the
``missing'' density in the middle of each spindle is due to the presence of $\nChr=20$
identical voids, we can infer the average void profile, which is well-approximated by a
circle arc (solid curve) with waist diameter $\sqrt{ab} = (1.0 \pm 0.1) \um$ and length
$\beta = (11 \pm 1) \um$ (both 95\% CI). To aid visualization of the physical shape of the
void, the fit is also shown reflected in the $x-$axis (dashed curve).  (g) The formation
of director-aligned, tactoid-like voids around compact inclusions can be seen in everyday
nematic materials, for example cylindrical lip gloss tubes inserted into a straw broom
head whose loose end is wrapped with a hair tie. A US quarter dollar coin is included for
scale. Error bars in (e) and (f) show SE. All scale bars $5 \um$.}
\end{figure}

\initializeSIsec{\metaphasePlateSIsec}
\ifnum \headings > 0
\section{Chromosomes Cause Tactoid-Like Voids in the Microtubule Network}
\fi
To understand the origins of these micron-scale density inhomogeneities and how they might
affect orientational fluctuations, we labeled chromosomes by expressing H2B-EGFP and
microtubules by SiR-tubulin staining, and used 3D confocal microscopy to image the
internal structure of living oocyte spindles (Methods 8 \& 9). Confocal micrographs show
that the microtubule network is a contiguous material perforated by voids (i.e. regions of
low microtubule density) surrounding each embedded chromosome
(Fig.~\ref{fig:voids}(a\&b)).  Consistent with recent results demonstrating that condensed
chromosomes are microtubule-impermeable~\cite{Schneider2022}, void cross-sections in the
metaphase plate (i.e.~the $x=0$ plane) closely follow chromosome boundaries, and may be
approximated as ellipses with long and short axes $a$ and $b$ respectively, where $a/b
\approx 2.5$ (S.I.~\metaphasePlateSIsec). In the $\xhat$-direction, the voids extend much
further than chromosomes, along most of the length of the spindle
(Fig.~\ref{fig:voids}(c)).

We were not able to use the micrographs to quantify individual void profiles along $\xhat$
because, for a significant portion of their lengths, void widths are smaller than the
microscope's resolution. Instead, we used LC-PolScope data to infer the average void
profile along $\xhat$. To do this, we fit the microtubule cross-sectional density
$\rho_{x}(x) = \langle \rho(\Rvec,t) \rangle_{y,z,t}$ as a function of position $x$ along
the spindle long axis (Fig.~\ref{fig:voids}~(d \& e); S.I.~3). The density profile $\rho_{x}(x)$
reaches a minimum at the metaphase plate and a maximum near the poles
(Fig.~\ref{fig:voids}(e)). By assuming all of this ``missing'' density in the central
spindle is due to voids, we infer that $\sim$10\% of the total volume, and $\sim$15
\% of the metaphase plate area, of MII spindles is taken up by voids. Assuming further
that all spindles contain $\nChr=20$ voids (one per chromosome), we find that average void
profiles along $\xhat$ are well-approximated by circle arcs with waist diameter around $1
\um$ (Fig.~\ref{fig:voids}~(f)), which may be interpreted as the geometric mean of the
long and short axes of the void's $x=0$ cross-section, $\sqrt{ab}$
(Fig.~\ref{fig:voids}(c, \textit{bottom left inset})). The simplest 3D shape that would
generate such a missing retardance profile is a tactoid, the form generated by revolving a
circle arc of chord length $\beta$ about its chord, which would have a circular
cross-section in the metaphase plate ($a=b$). Since the voids in MII spindles have
non-circular $x=0$ cross-sections ($a \neq b$), their 3D shapes are not true tactoids,
but rather are ``tactoid-like'' in the sense that their average profile along $\xhat$ is a
circle arc. Tactoid-shaped holes have been observed previously in both synthetic and
biological nematics, and are known as ``negative tactoids'' or
``atactoids''~\cite{Bernal1941, Nastishin2005, Almohammadi2023}; as in spindles, these
strutures tend to spontaneously align with the nematic director (\cite{Yi2013, Zhou2014,
Genkin2018}, Fig.~\ref{fig:voids}(g)).

\ifnum \headings > 0
\section{Chromosomes/Voids Are Locally Ordered in the Spindle Mid-Plane}
\fi
\initializeSIsec{\simsSIsec}
\initializeSIsec{\metaphaseModelSIsec}

\begin{figure}[t!]
  \centering
  \includegraphics[width=0.7\textwidth]{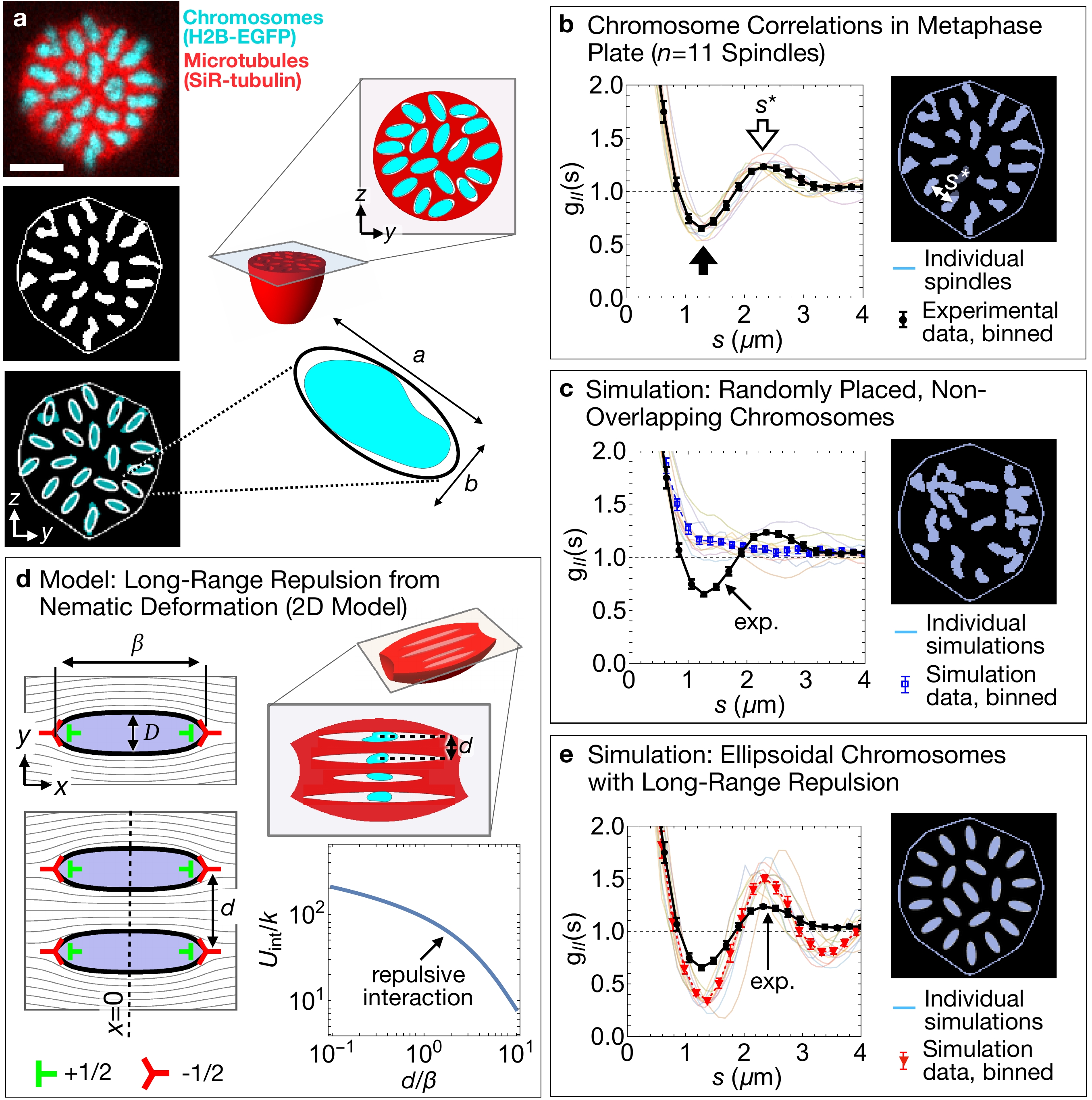}
 \caption{\label{fig:chromosomeRepulsion2} Metaphase plate configurations in MII spindles
are consistent with a model in which chromosomes repel each other over distances of
several microns, but inconsistent with random positioning of non-overlapping
chromosomes. (a) \textit{Top:} Confocal micrograph of the metaphase plate ($x=0$ plane) of
a living MII spindle, labeled as in Fig.~\ref{fig:voids}. \textit{Middle:} Binarization
identifies $\nChr = 20$ chromosome sections, with centers spaced $\sim 2\um$
apart. Spindle boundary in white. \textit{Bottom:} Chromosome sections approximated by 20
identical ellipsoids. In this spindle, $a = 2.1 \um$ and $b = 0.8 \um$.  Scale bar $5
\um$. (b) \textit{Left:} At separations $s$ significantly less than the typical chromosome size ($s \ll b$), the
correlation function $g_{II}(s)$ is much greater than one (dashed black line). At larger
separations, $g_{II}(s)$ first reaches a local minimum at $(1.27 \pm 0.04) \um$ (black
arrow) and a local maximum at $s^{*} = (2.4 \pm 0.1) \um$ (white arrow). \textit{Right:}
The separation $s^{*}$ corresponds to the distance away from each chromosome where it is
most probable to find another chromosome. (c) \textit{Left:} Pair correlation functions
for simulated data of non-overlapping, randomly placed 2D chromosome sections lack local
extrema, and decay monotonically to 1.  Each colored curve is obtained from a simulation
initialized with one experimental data set. \textit{Right:} Snapshot of one simulation,
initialized with the spindle boundary and binarized chromosome sections shown in (a).  (d)
Model for deformation-induced long-range repulsion between negative-tactoid-like
inclusions in a 2D nematic. In this model, the orientation field (thin gray curves)
everywhere in the $z=0$ plane is fully specified by two rows of $\pm 1/2$ topological
defects; each row is parallel to $\xhat$ and their centers lie along $x=0$ (dashed black
line). Inclusion boundaries (thick black curves) are those integral curves of the nematic
field that pass through both $- 1/2$ defects in a given row
(S.I.~\metaphaseModelSIsec). Defect spacings within a row are determined by the geometric
parameters $\beta$ and $D$, chosen to match the experimentally observed void length and
waist curvature (Fig.~\ref{fig:voids} (c \& f)). The interaction potential
$U_{\text{int}}$, normalized by elastic constant $k$, induced by deformation of the
nematic field is repulsive at all center-to-center distances $d$. (e) \textit{Left:}
Correlation functions for simulated data of 20 ellipsoids interacting in the $x=0$ plane
via a repulsive $U(s) \propto s^{-6}$ potential display a local minimum [maximum] at
$(1.32 \pm 0.04) \um$ [$(2.4 \pm 0.1) \um$]. \textit{Right:} Snapshot of one simulation
performed with spindle boundary and chromosome shape parameters from (a)
(S.I.~\simsSIsec). Error bars in all plots indicate SE.}
\end{figure}

To investigate the configuration of voids within the metaphase plate, we took advantage of
the fact that, in the $x=0$ plane, chromosome cross-sections provide a good proxy for void
cross-sections (Fig.~\ref{fig:voids}(b)), but are easier to identify since chromosomes are
high-contrast, isolated, compact objects. To study chromosome cross-section
configurations, we binarized images of the metaphase plates of several spindles
(Fig.~\ref{fig:chromosomeRepulsion2}(a)). To detect correlations between chromosome
positions, we use the pair correlation function $g_{II}(s)$, which quantifies the average
correlation between the pixel value $I$ at pairs of points separated by a distance $s$ in
the metaphase plate (S.I.~\metaphasePlateSIsec).  At separations much less than the
smaller chromosome dimension ($s \ll b \approx 1 \um$), $g_{II}(s) \gg 1 $; this reflects
the fact that a white pixel is very likely to immediately neighbor other white pixels.  At
larger separations, we observe a local minimum at $(1.27 \pm 0.04) \um$ corresponding to
the the presence of a ring around each chromosome that is depleted of other chromosomes,
and a local maximum at $s^{*} = (2.4 \pm 0.1) \um$ indicating a ring enriched in
chromosomes (Fig.~\ref{fig:chromosomeRepulsion2}(b), black and white arrows; uncertainty
given by empirical bootstrapping). To confirm this interpretation, we ran a Monte Carlo
simulation that takes as inputs the experimentally determined set of binarized chromosome
sections, and re-arranges them into a random configuration (S.I.~\simsSIsec). Random
configurations appear strikingly different to the experimentally observed ones, and the
corresponding $g_{II}(s)$ lacks local maxima and minima
(Fig.~\ref{fig:chromosomeRepulsion2}(c)), indicating that these simulations lack the local
order (i.e.~spatial anti-correlation) observed in the experimental data.  Local ordering
of chromosomes and their associated voids also explains the previously noted peak in
$s_{\nn}(q_{0},q_{y})$ (Fig. 2(c)), since large fluctuations concentrated in
regularly-spaced voids cause a peak in the orientational correlation function
(S.I.~6). Consistent with this interpretation, the characteristic spacing between voids,
$s^{*}= (2.4 \pm 0.1) \um$, agrees, within experimental uncertainty, with the position of
the correlation function peak, $\lambda^{*}= (2.2 \pm 0.1) \um$.

\ifnum \headings > 0
\section{Locally Ordered Chromosome Configurations Are Consistent with Long-Range
  Repulsion from Deformation of the Nematic Field}
\fi
We next turned to the origin of the forces responsible for chromosome ordering in the
metaphase plate. In liquid crystal physics, it is well-known that deformation of the
nematic field around micron-size inclusions can create long-range forces that cause the
inclusions to self-organize into structured arrays~\cite{Poulin1997, Sharma2014}. We
therefore hypothesized that, in MII spindles, a similar force might cause the regularly
spaced chromosome configurations we observe. To explore this effect, we constructed an
analytically tractable 2D model in which the void surrounding a chromosome is represented
as a topological quadrupole made up of two $+1/2$ and two $-1/2$ defects (\cite{Vafa2020};
Fig.~\ref{fig:chromosomeRepulsion2}(d, \textit{top left})). In this model, void boundaries
are defined as those integral curves of the director that pass through the outer pair of
$-1/2$ defects. With this construction, the length $\beta$ and width $D$ of the void
uniquely determine the spacing of the defects within a row and thus, for an isolated void,
the orientation field everywhere in space (S.I.~\metaphaseModelSIsec). For a pair of
parallel, director-aligned voids whose centers lie along a line perpendicular to the
far-field director $\xhat$, the deformation-induced interaction potential
$U_{\text{int}}(d)$ decays monotonically for all values of $\beta, D$, and
center-to-center separation $d$, implying the existence of a repulsive force between 2D
voids, independent of the details of void geometry (Fig.~\ref{fig:chromosomeRepulsion2}(d);
S.I.~\metaphaseModelSIsec).

To investigate whether long-range repulsion between 3D voids can account for the observed
metaphase plate configurations, we performed a series of simulations where chromosome/void
sections are represented by ellipses interacting via a long-ranged repulsive potential
(S.I.~\simsSIsec). In each simulation, the ellipse geometry and metaphase plate boundary
are determined from a specific experimental data set. Since we do not know the form of the
interactions between non-axisymmetric voids in 3D, we repeat our simulations using four
different long-range repulsive potentials, and find that the specific form of the
interaction does not significantly affect the final chromosome configuration: all
simulations produce configurations similar to the experimentally observed one, with
features such as an outer ring of $\sim$15 mostly radially oriented chromosomes/ellipses,
and local extrema of $g_{II}(s)$ near 1.3 and $2.4\um$
(Fig.~\ref{fig:chromosomeRepulsion2}(e), S.I.~\simsSIsec). Thus, our observations of local
ordering of chromosomes in the metaphase plate are consistent with the presence of
long-range repulsion arising from deformation of the nematic field.

\ifnum \headings > 0
\section{Conclusions}
\fi
In this work, we have demonstrated that chromosomes are locally ordered in the metaphase
plate of MII mouse oocytes, implying the existence of repulsive interchromsomal forces
acting perpendicular to the spindle axis (Fig.~\ref{fig:chromosomeRepulsion2}). The
micron-scale distances between chromosome surfaces (S.I.~\simsSIsec) are far larger than
can be accounted for by known forces, such as those arising from electrostatic
repulsion~\cite{Kesler2020} or steric interactions between chromosome-associated
proteins~\cite{Cuylen2019}. To explain this observation, we proposed a novel mechanism
whereby distortion of the microtubule network around chromosomes causes repulsion between
them. Our model relies on the key assumption that stress and torque propagate through the
microtubule network according to the predictions of nematic elasticity, i.e.~that the
microtubule network has the mechanical properties of a nematic liquid crystal. This
assumption is consistent with several other observations of mouse oocyte spindles, in
particular the shape of the spindle boundary (Fig.~\ref{fig:polscope}(b-f)), the
steady-state pattern of microtubule orientation in the spindle interior
(Fig.~\ref{fig:polscope}(g\&h)), the functional form of spatial correlations of
orientational fluctuations (Fig.~\ref{fig:flucs}(c)), and the appearance of
director-aligned, tactoid-like voids around embedded chromosomes
(Fig.~\ref{fig:voids}). Consistent with our findings, previous work in \textit{Xenopus}
egg extract~\cite{Brugues2014, Oriola2020} and human tissue culture
cells~\cite{Conway2022} showed that nematic models accurately predict several aspects of
spindle structure and dynamics in those systems also. Taken together, these results
suggest that nematic elasticity plays a fundamental role in determining both spindle
structure and chromosome organization in large spindles across organisms and cell types.


\newpage 
\section{References}
\bibliographystyle{unsrt}
\bibliography{/Users/colmkelleher/Documents/latex_harvard/orientation_paper_draft/library.bib}
\newpage

\newpage

\newpage

\newpage

\newpage


\end{document}


\title{Supplemental Information for ``Long-Range Repulsion Between Chromosomes in Mammalian Oocyte Spindles ''}
\date{\today}

\maketitle

\tableofcontents
\newpage

\section{Formation and Lifetimes of MI \& MII Spindles in Mouse Oocytes}
\begin{figure}[t!]
  \centering
  \includegraphics[width=\textwidth]{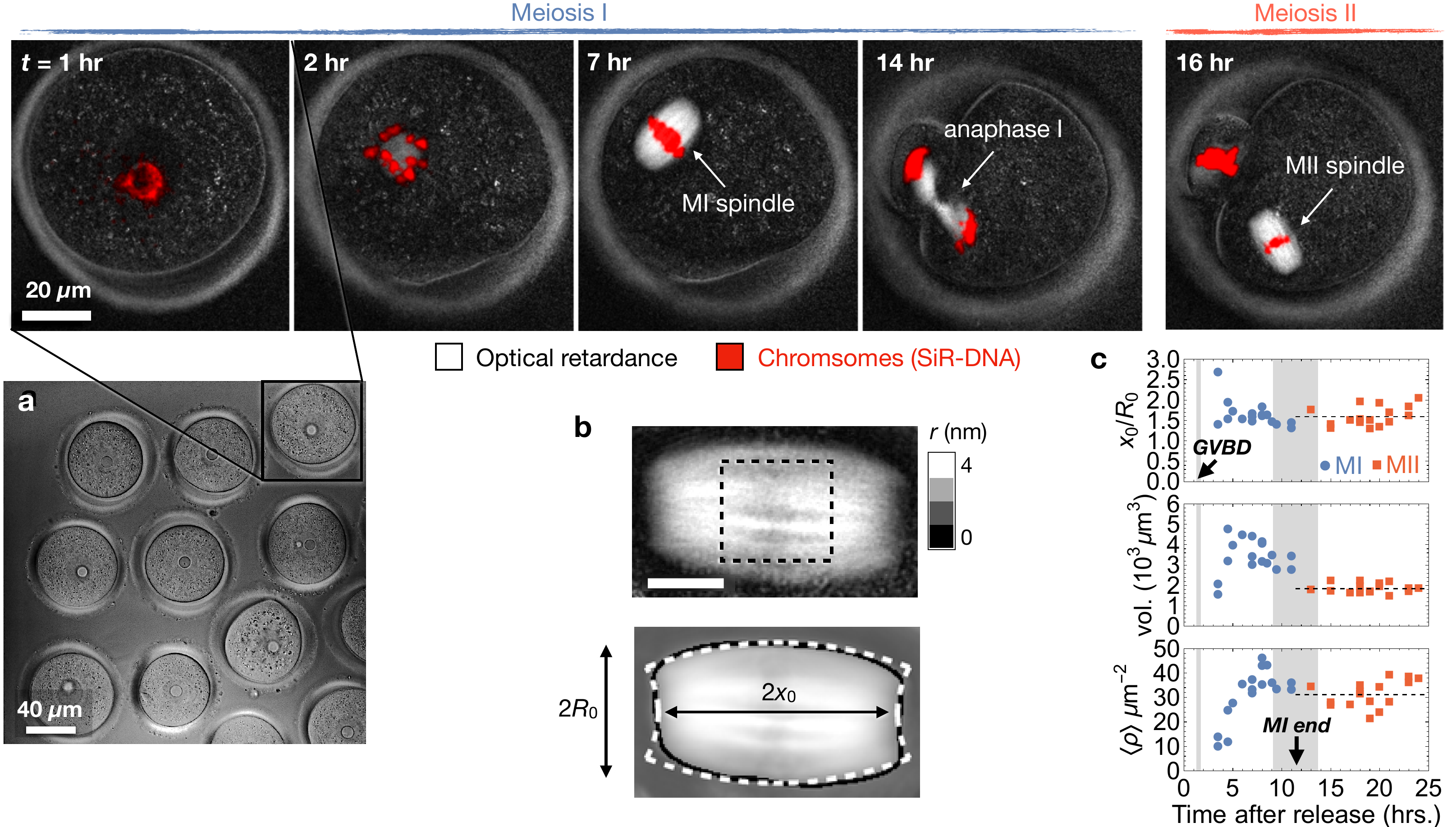}
  \caption{\label{fig:mIandMII} With respect to the parameters measured by the
LC-PolScope, MII oocytes remain in steady state for at least 12 hours after completion of
meiosis I.  (a) \textit{Main Panel:} GV-arrested oocytes. \textit{Zoom:} Time series of
maturation of an individual oocyte, showing the formation of the first and second (MI and
MII) meiotic spindles, as well as anaphase I. Times are given relative to release from
meiotic arrest (Methods 1). Spindles are imaged using LC-PolScope (grayscale); chromosomes
are stained with SiR-DNA and imaged using epifluorescence (red). (b) For both MI and MII
spindles, length $2 x_{0}$ and maximum width $2 R_{0}$ are measured from the best-fit
pole-indented tactoid (Methods 3). (c) Plots showing the time-course of the spindle's
aspect ratio $x_{0}/R_{0}$, volume, and microtubule cross-section density $\langle
\rho\rangle$ (Eqn.~\ref{e1:retToDens1}). Release of spindles from meiotic arrest occurs at
$t=0$. Grey bars indicate the distributions of times required to complete GV breakdown and
times required to complete meiosis I (mean $\pm$ standard deviation; $n= 71$
oocytes). Density is calculated using the retardance measured in an $8 \um \times 8 \um$
box at the center of each spindle (dashed black square in (b)), and sample thickness equal
to the spindle diameter, $T=2 R_{0}$. For MI spindles (blue circles), density and volume
grow from zero over the course of several hours. For MII spindles (orange squares), all parameters reach
steady-state $ \lesssim $ 1 hr after completion of
meiosis I, and no trend is evident for at least 12 hrs thereafter, and are consistent
with Gaussian distribution about the mean (black dashed line, Kolmogorov Smirnov test with $p > 0.05$).}
\end{figure}

As described in Methods 1, we extract germinal-vesicle-stage (GV-stage) oocytes from mouse
ovaries and mature them \textit{in vitro}. LC-PolScope observations indicate that, after
completion of anaphase I, MII oocytes spindles remain in steady state for at least 12 hours
(S.I.~Fig.~\ref{fig:mIandMII}).
\newpage

\section{Retardance Images Behave As If Light Were Collimated as It Passes Through Spindle}
\label{sec:collimated}
\begin{figure}[t!]
  \centering
  \includegraphics[width=\textwidth]{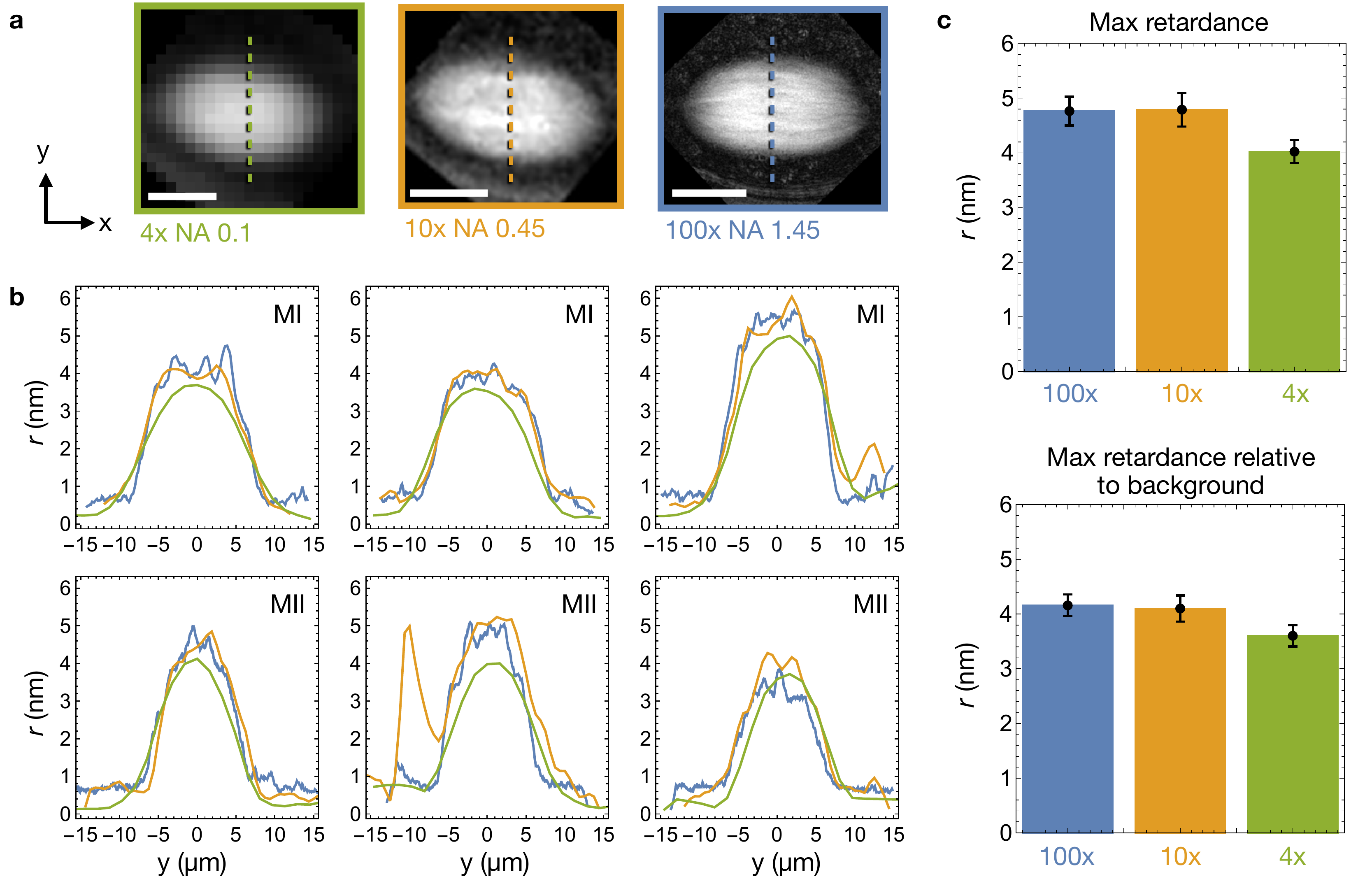}
  \caption{\label{fig:objectives} LC-PolScope retardance profile of spindles does not
depend on NA of optical system, consistent with collimated light assumption. (a) The same
MI spindle imaged with three different objective lenses: 4x (NA 0.3); 10 x (NA 0.7); 100x
(NA 1.4). Scale bars $10 \um$. (b) Retardance profiles through the center of three MI
spindles (\textit{top row}) and three MII spindles (\textit{bottom row}), in the direction
perpendicular to the spindle axis (dashed lines in (a)). (c) Histograms of
the maximum retardance in the three MI line profiles shown in (b). The maximum measured
retardance is similar in all cases, as is the maximum retardance with the background
retardance values subtracted. Error bars indicate SE.}
\end{figure}

In this work, we treat all LC-PolScope data as if it represented a simple projection over
the optical axis $\ohat$ (Main Text Eqns.~1). This is equivalent to modeling the light
passing through the spindle as collimated, i.e.~we assume that the optical section depth
of the LC-PolScope is much larger than the spindle diameter. This assumption is consistent
with previous measurements of invertebrate oocytes in a quantitative polarization
microscopy system similar to the LC-PolScope~\cite{Sato1975}. In that work, it was shown
that measured retardance values do not depend on the details of the optical train, in
particular on the numerical apertures (NAs) of the objective and condenser lenses,
consistent with the assumption of collimated light. However, we used immersion objectives
with significantly higher NAs than reference~\cite{Sato1975}, and, to our knowledge, there
is not a simple way to theoretically estimate the optical section depth of
``semi-coherent'' microscopy like LC-PolScope~\cite{Tran2022}. To experimentally check
whether our results depend on the details of the optical train, we measured the retardance
profiles of MI and MII spindles with three different objectives, whose NAs varied from
0.10 (magnification 4x) to 1.45 (magnification 100x). Consistent with the results of the
previous study, we found that our retardance measurements do not depend strongly on the NA
or magnification of the objective lens used, and conclude that, in all cases, transmitted
light behaves as though it were approximately collimated
(S.I.~Fig.~\ref{fig:objectives}). In all LC-PolScope experiments, we used a condenser lens
with NA 0.52, with the condenser iris in the half-closed position such that the effective
condenser NA $\approx 0.25$. \newpage

\section{Calculation of Microtubule Cross-Sectional Density from
  Observed Retardance}
\label{sec:densAndRet}
\subsection{Retardance of Microtubule Arrays Aligned Perpendicular to the Optical Axis}
Microtubules exhibit ``form birefringence''; that is, their birefringence arises
predominantly from their rod-like shape and their high refractive index relative to the
surrounding medium, rather than ``stress birefringence'' (arising from anisotropic
mechanical strain) or ``intrinsic birefringence'' (arising from anisotropic distribution
of oriented chemical bonds)~\cite{Sato1975}. Here, we construct a model, based on Wiener's
equation for form birefringence~\cite{Wiener1912}, that relates the observed retardance of
a sample of microtubules coherently aligned in a direction $\xhat$ perpendicular to the
optical axis $\ohat$ with the average 2D cross-sectional density $\rho_{0}$ of the
microtubules in the plane perpendicular to $\xhat$. By definition, the retardance of a
sample is proportional to its thickness $T$ and the refractive index mismatch $\Delta n =
n_{\text{slow}}-n_{\text{fast}}$ between light polarized along the fast and slow optical
axes,
\begin{equation*}
  r = T \Delta n.
  \end{equation*}
For materials like microtubules, whose birefringence arises from their rod-like shape,
$\Delta n$ is related to the volume fraction $f$ and refractive index $\nMT$ of the
microtubules, as well as the refractive index $\nC$ of the surrounding cytoplasm,
\begin{equation*}
\Delta n= f \frac{(\nC^{2}-\nMT^{2})^{2}}{2 \nC (\nC^{2}+\nMT^{2})}.
  \end{equation*}
This expression assumes $f \ll 1$; this assumption is generally appropriate for spindle
microtubules~\cite{Nixon2017}. For a bulk material composed of aligned rods, it is the
also case that
\begin{equation*}
f = \pi \Big{(} \frac{d_{MT}}{2} \Big{)}^{2} \rho_{0},
  \end{equation*}
where $d_{MT}$ is the diameter of a single rod. Putting these equations together, we
obtain an equation relating $r$ and $\rho_{0} $,
\begin{SIequation}
  \label{e1:retToDens1}
 \rho_{0}  = \frac{r}{T A_{0}},
\end{SIequation}
where the so-called retardance area $A_{0}$ characterizes the contribution of a single
microtubule to the measured retardance,
\begin{equation*}
A_{0}\equiv \Big{(}\frac{\pi d_{MT}^{2}}{4}\Big{)}^{2}\frac{(\nC^{2}-\nMT^{2})^{2}}{2 \nC (\nC^{2}+\nMT^{2})}.
\end{equation*}
The microtubule-related quantities $d_{MT}$ and $\nMT$ have been previously measured, and
should not depend strongly on cell type: $d_{MT} \approx 24$ nm and $\nMT \approx
1.512$~\cite{Olmsted1973, Sato1975}. To our knowledge, the refractive index of mouse
oocyte cytoplasm has not been measured. For other metazoan cell types, $\nC$ typically
lies in the range 1.35-1.40~\cite{Sato1975, Gul2022}. Assuming that $\nC = 1.375 \pm 0.025$, we obtain
$A_{0} =( 6.2 \pm 1.8)$ nm$^{2}$, consistent with previous experimental
measurements of this quantity \textit{in vitro}, which yield $A_{0} \approx 7.5 $
nm.~\cite{Oldenbourg1998}. Using the measured values of microtubule cross-sectional
density given in the Main Text, we find $f \approx 0.04$, consistent with the assumption
$f \ll 1$ used in the above derivation.

\subsection{Average Microtubule Cross-Sectional Density in MI and MII Spindles}
To estimate the average microtubule cross-sectional density $\langle \rho \rangle$
(S.I.~Fig.~\ref{fig:mIandMII}(c)), we first calculate the average retardance in a $8 \um \times
8 \um$ box in the center of the spindle, and take the spindle diameter as the sample
thickness, $T \approx 2 R_{0}$, and directly apply Eqn.~\ref{e1:retToDens1}.

\subsection{Predicted Retardance Profile of Oocyte Spindle}
\begin{figure}[t!]
  \centering
  \includegraphics[width=0.7\textwidth]{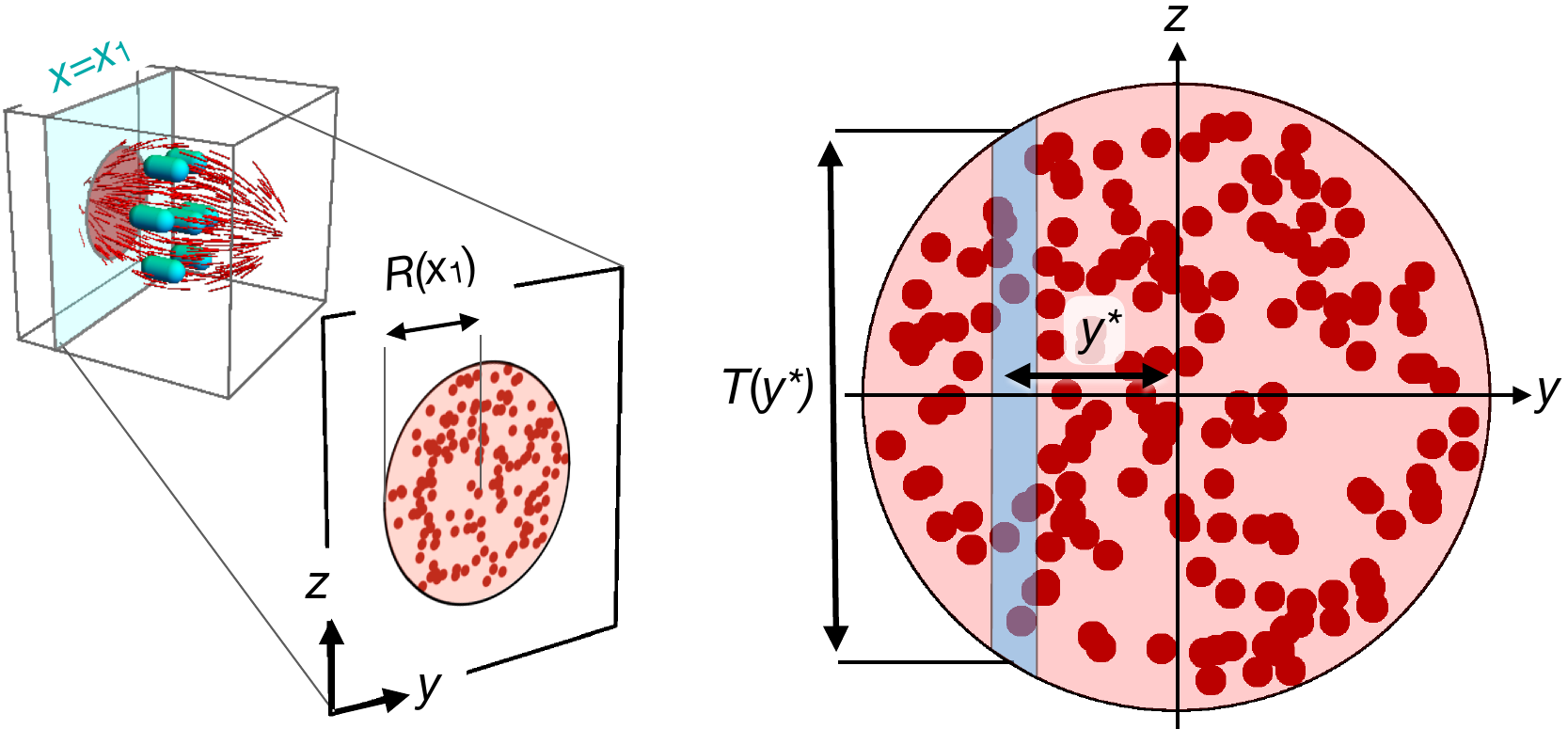}
  \caption{ \label{fig:spidleRetProfile} Thickness profile of spindle with long axis $\xhat$
    perpendicular to optical axis $\zhat$. In the plane $x=x_{1}$, the spindle cross-section
    has radius $R(x_{1})$, and the spindle boundary is described by the equation
    $y^{2}+z^{2} = R(x_{1})^{2}$. The optical path length $T$ through the spindle
   along the ray $y=y^{*}$ is given by $T(y^{*}) = 2 (R(x_{1})^{2}-(y^{*})^{2})^{1/2}$.}
\end{figure}

To measure the dependence of microtubule density on position along the spindle axis (Main
Text Fig.~3 (d \& e)), we construct a more detailed model which takes into account the
fact that spindles do not have uniform thickness, but are well-described as surfaces of
revolution with axes of revolution $\xhat$,
\begin{equation*}
\Rvec(x,\phi) = \{ x, R(x) \cos{\phi}, R(x) \sin{\phi} \},  
\end{equation*}
where $R(x)$ is the spindle radius at postion $x$ along its long axis, and $\tan{\phi}
\equiv y/z$ is the cylindrical polar coordinate relative to the $\xhat$-axis. Assuming collimated
light and that the spindle is centered at the origin, the sample thickness corresponding
to a specific position $\{x,y \}$ in the LC-PolScope image plane is given by the formula
$T(x,y) = 2(R(x)^{2}-y^{2})^{1/2}$, and Eqn.~\ref{e1:retToDens1} may be rewritten as
\begin{SIequation}
  \label{sieq:retProfile}
r(x,y)  = 2 \langle \rho \rangle_{z}(R(x)^{2}-y^{2})^{1/2},
\end{SIequation}
where $\langle \rho \rangle_{z}$ is the cross-sectional density averaged over the optical
path through the spindle (S.I.~Fig.~\ref{fig:spidleRetProfile}). To find the average
cross-sectional density in the retardance slice $x=x_{i}$, we treat density
$\rho_{x}(x_{i})$ and spindle radius $R(x_{i})$ as independent fit parameters in the
equation
\begin{equation*}
r(x_{i},y)  = 2 \rho_{x}(x_{i})(R(x_{i})^{2}-y^{2})^{1/2},
\end{equation*}
where the density profile $\rho_{x}(x)$ along the spindle axis may be interpreted as the
3D density profile averaged over $y$ and $z$, $\rho_{x}(x) \equiv \langle \rho (x,y,z)
\rangle_{y,z}$. For time-averaged retardance images, we consider $\rho_{x}(x)$ to be
averaged over time as well, $\rho_{x}(x) \equiv \langle \rho (x,y,z,t) \rangle_{y,z,t}$.

\newpage

\section{Justification for Fitting Spindle Boundary to Pole-Indented Tactoid; Comparison of
  Shape Parameters from Confocal Slices with Those from LC-PolScope Retardance Images}
\begin{figure}[t!]
  \centering
  \includegraphics[width=\textwidth]{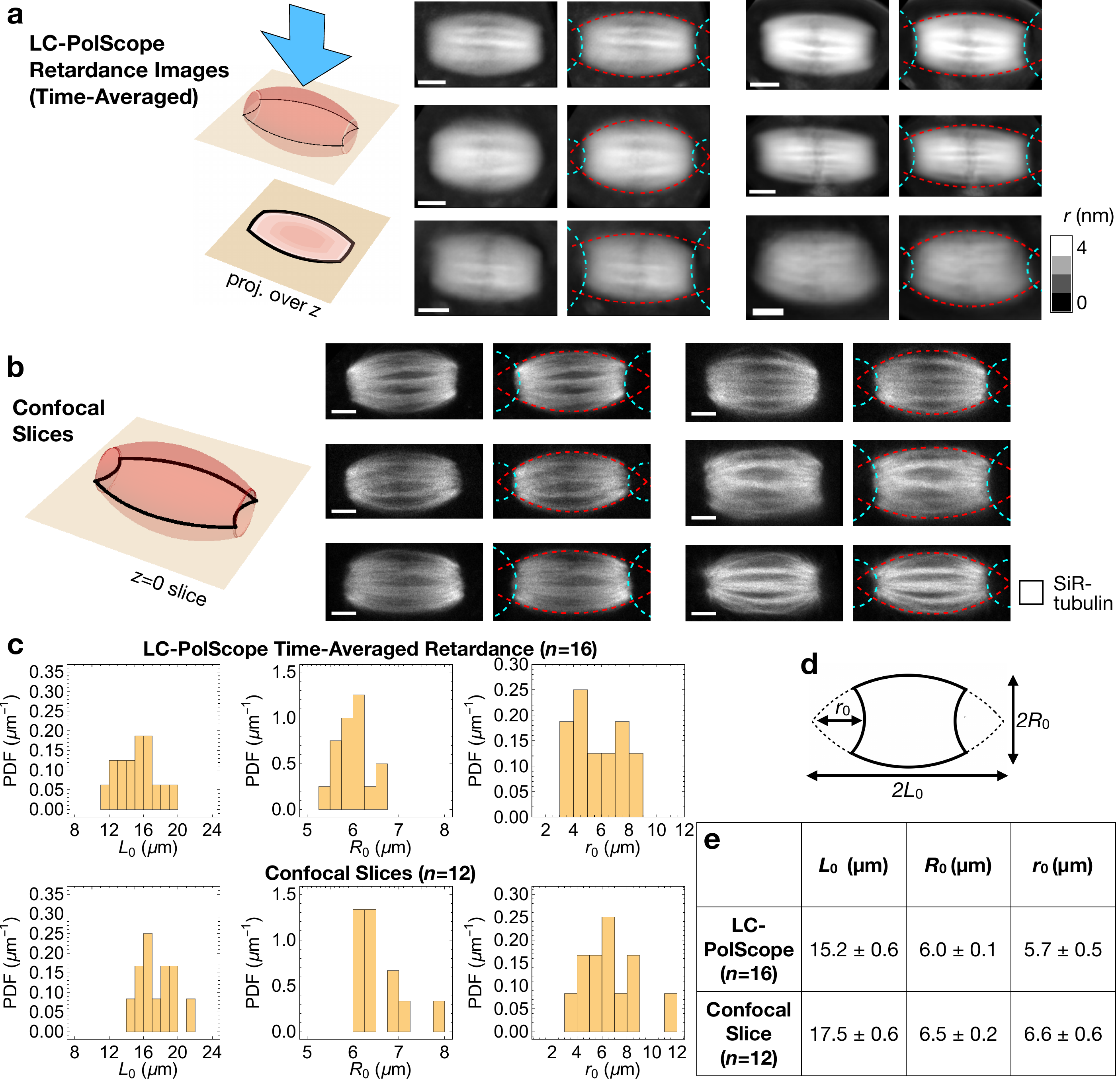}
  \caption{ \label{fig:horiz-sections} Spindle shape from confocal slices vs spindle shape
from LC-PolScope. (a) LC-PolScope time-averaged retardance images $\langle r \rangle_{t}$
(\textit{left panels of image pair}), which we interpret as $z$-projections of microtubule
cross-sectional density, and the corresponding best fit circle arc segments (\textit{right
panels}, dashed colored lines). (b) Slices of confocal stacks that contain the spindle's
central axis (\textit{left panels}), and the corresponding best fit circle arc segments
(\textit{right panels}, dashed colored lines). (c) Histograms of best-fit shape parameters
for $n=16$ LC-PolScope images and $n=12$ confocal slices. (d) The generatrix of a
pole-indented tactoid is made up of the four circle arc segments shown, which are fully
determined by the three shape paramaters $L_{0}$, $R_{0}$, and $r_{0}$.  (e) The shape
parameters found by fitting LC-PolScope retardance images are smaller than those found by
fitting confocal slices ($p < 0.01$ for $L_{0}$ and $R_{0}$; Student's t-test). All scale bars $5 \um$. }
\end{figure}

In the analysis presented in Fig.~1 of the Main Text, we fitted empirically observed
spindle boundaries, obtained from LC-PolScope retardance images, to the generatrix of a 3D
shape we call a pole-indented tactoid (Fig.~1 (c \& d);
S.I.~Fig.~\ref{fig:horiz-sections}(a)). We chose this shape for two reasons: first, a
very similar shape has been shown to arise from a mechanical model in which
molecular-motor-driven contractile forces balance stresses from nematic elasticity at the
spindle boundary~\cite{Oriola2020}. Second, the generatrix of a pole-indented tactoid
(Fig.~1 (d)) is a good fit to the boundaries of cross-sections of the spindle that include
its central axis, as measured by confocal microscopy (S.I.~Fig.~\ref{fig:horiz-sections}
(b)). The shape parameters ($L_{0}$, $R_{0}$, and $r_{0}$) found by fitting confocal
sections are systematically $\sim$10\% greater than those found from analysis of
LC-PolScope retardance images (S.I.~Fig.~\ref{fig:horiz-sections} (c \& d)), presumably
due to errors associated with the fact that the long axes of spindles imaged with the
LC-PolScope are not perfectly perpendicular to the optical axis (Methods 2) and errors in
identifying the spindle boundary from a retardance images (Methods 3). We consider the
parameters obtained from fitting the confocal images to more accurately characterize the
true shapes of living MII spindles.
\newpage

\section{2D Circle Arcs Model Is a Good Approximation of
  $\zhat$-Projected 3D Circle Arcs Model}
\begin{figure}[t!]
  \centering
  \includegraphics[width=\textwidth]{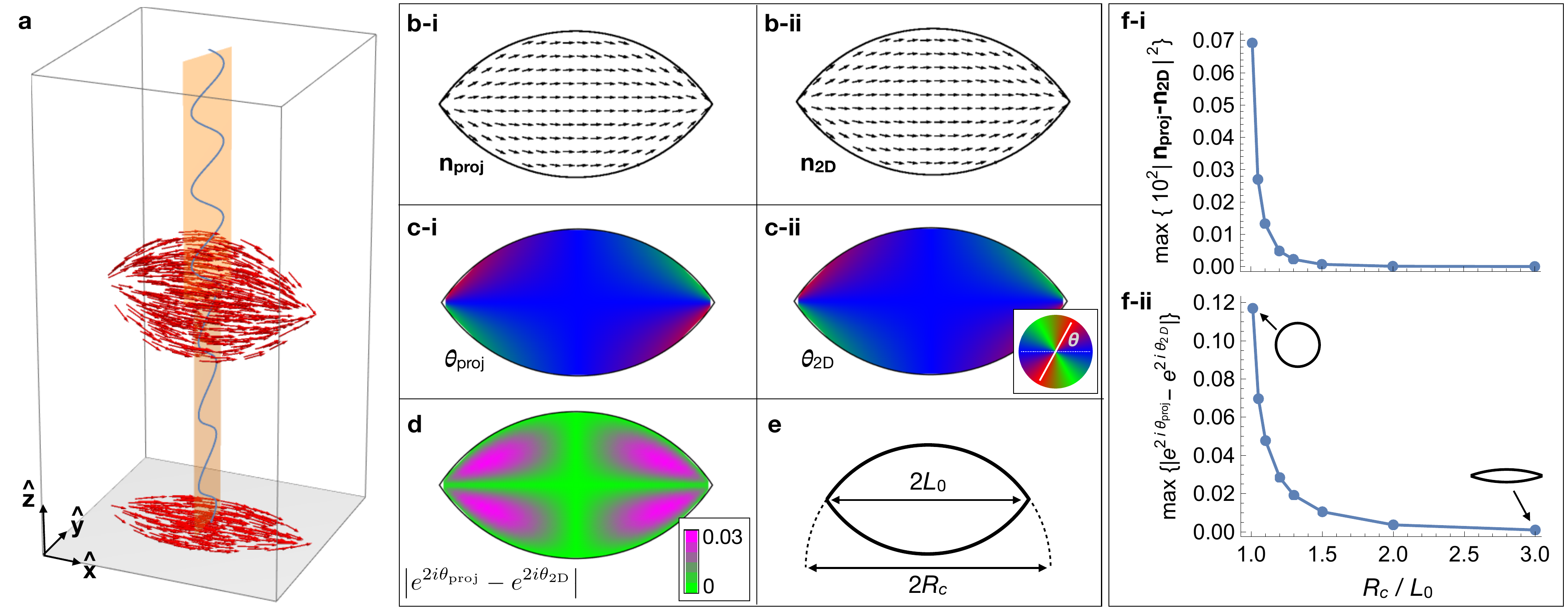}
  \caption{ \label{fig:errorFromProj} Comparison of 3D nematic fields projected into 2D,
and 2D nematic fields obtained from a 2D model. (a) At each pixel of a 2D image, the
LC-PolScope measures the nematic field averaged over the thickness of the spindle; (b) The
nematic field obtained by (i) assuming a 3D circle arcs model, integrating over the
optical axis $\zhat$, and normalizing the resulting field, and (ii) a 2D version of the
circle arcs model. (c) Angles obtained from the fields in (b). (d) Absolute difference in
2D nematic order parameter between the fields in (c). (e) All data in this panel is
generated from a model in which spindle geometry fully is characterized by the two
parameters $R_{c}$ and $L_{0}$. In terms of the parameters of the pole-indented tactoid
(S.I.~Fig.~\ref{fig:horiz-sections}), $R_{c} = (L_{0}^{2}+R_{0}^{2})/2R_{0}$. (f) The
maximum absolute difference between the fields in (b) and the fields in (c) and is
greatest for the smallest values of the spindle aspect ratio $R_{c}/L_{0}$. In oocyte
spindles, $R_{c}/L_{0} \approx 1.5$ (S.I.~Fig.~\ref{fig:horiz-sections}), and the maximum
error between the 2D and 3D-projected nematic fields is about 1\%.}
\end{figure}

The orientation fields we obtain from LC-PolScope are not the full 3D orientation fields;
rather, they are those fields projected over an axis, $\zhat$, that is perpendicular to
the spindle long axis $\xhat$ (Main Text Fig.~1). Therefore, previously derived
predictions for 3D nematic tactoids, e.g.~the circle arcs model~\cite{Williams1986}, do
not directly apply to our data (S.I.~Fig.~\ref{fig:errorFromProj}). Here, we estimate the
error incurred by approximating the angle fields derived from projections of 3D circle
arcs with the the 2D circle arcs model. To do this, we construct the 3D nematic field for
the nematic tactoid that is uniquely specified by the defect half-spacing $L_{0}$ and the
radius of curvature $R_{c}$ of the tactoid boundary
(S.I.~Fig.~\ref{fig:errorFromProj}(e);~\cite{Prinsen2003}). We then calculate the
projection of that nematic field into 2D along an axis perpendicular to the tactoid long
axis, and compare it to the field generated by the 2D family of circle arcs generated with
the same defect spacing and radius of curvature. Our comparison includes both the fields
in the central region of the tactoid (the region that describes oocyte spindles very well; Main Text
Fig.~1), as well as fields in the polar regions (absent in oocyte spindles).  The error we
calculate here can therefore be considered as an upper bound on the error associated with
comparing the measured $\zhat$-projection of the 3D orientation field with a 2D circle
arcs model in mouse oocyte spindles. In the geometry relevant to mouse oocyte spindles
($R_{c}/L_{0} \approx 1.5$), the error introduced by using a 2D circle arcs model to
describe projected 3D data is around 1\% (S.I.~Fig.~\ref{fig:errorFromProj}(f)).

\newcommand{\nihat}{ \hat{N}_{i}}
\newcommand{\njhat}{ \hat{N}_{j}}
\newcommand{\dNj}{ \delta N_{j}}
\newcommand{\dNy}{ \delta N_{y}}
\newcommand{\dNz}{ \delta N_{z}}

\newpage

\section{Orientational Correlation Functions}
\subsection{Active Nematic Model for Equation of Motion for $\dN$ (Real \& Fourier Space)}
\label{subsec:nematicModel}
In equilibrium field theories, the equipartition theorem can be applied to an
appropriately diagonalized expression for free energy to calculate the spectrum of
fluctuations~\cite{deGennes1993}. In non-equilibrium systems like the spindle, we can
recover an analogous result by constructing an equation of motion and applying the
principles of fluctuating hydrodynamics~\cite{Zarate2006}. Quite generally, the director
field of an active nematic obeys an equation of motion derived from local torque
balance~\cite{Edwards2009, Brugues2014}
\begin{SIequation}
\gamma \big{(} \frac{\del \nihat}{\del t} + v_{j} \nabla_{j} \nihat -
  M_{ij} \njhat \big{)} = - h_{i} + \gamma S_{i}.
\label{eq:eqnOfMotion}
\end{SIequation}
Here, $\mathbf{v}$ is the hydrodynamic velocity field of microtubules plus surrounding
fluid, the tensor $\underline{\underline{M}} = 1/2 (\nabla \mathbf{v} - (\nabla
\mathbf{v})^T) + \lambda/2 (\nabla \mathbf{v} + (\nabla \mathbf{v})^T)$ describes the flow
alignment of microtubules, $\gamma$ is a rotational viscosity, $\lambda$ dictates how
microtubules orient in shear flows, $S_{i}$ is a noise term whose form we will specify
later. Using the one-constant approximation $k \equiv k_{\text{bend}} = k_{\text{twist}} =
k_{\text{splay}}$, the molecular field $h_{i}$ is given by $h_{i} = k \nabla^{2}
n_{i}$~\cite{deGennes1993}. In general, Eqns.~\ref{eq:eqnOfMotion} must be supplemented
by the constraint that the director is a unit vector, $|\mathbf{\hat{N}}| = 1$, and a
stress-balance equation which sets $\mathbf{v}$. For an active nematic at low Reynolds
number, the stress balance equation may be written
\begin{equation*}
  \underline{\underline{\sigma}}^{r} + \underline{\underline{\sigma}}^{v} + \underline{\underline{\sigma}}^{a} = 0, 
  \end{equation*}
  where $\underline{\underline{\sigma}}^{r}$, $\underline{\underline{\sigma}}^{v}$, and
$\underline{\underline{\sigma}}^{a}$ are the reactive, viscous, and active contributions to
the total stress, given by
\begin{align*}
   \sigma_{ij}^{r} &= p\,  \delta_{ij} - \frac{\lambda}{2} (n_{i}h_{i} + n_{j}h_{j});\\
     \sigma_{ij}^{v} & = -\eta \, \partial_{j}(\partial_{i}v_{j} + \partial_{j}v_{i});\\
     \sigma_{ij}^{a} & = -\zeta n_{i} n_{j},\\             
  \end{align*}
where $p$ is the hydrostatic pressure, $\eta$ is the dynamic viscosity of the composite
fluid, and $\zeta$ is the active stress density. However, it has previously been argued
~\cite{Brugues2014} that the flows in steady-state \textit{in vitro}
assembled Xenopus egg extract spindles are so small that the terms involving $\mathbf{v}$
in Eqn.~\ref{eq:eqnOfMotion} may be neglected. In that case, the equation of motion for
small fluctuations $\delta \mathbf{\hat{N}}$ about the director $\xhat$ becomes the
diffusion equation with non-conservative noise,
\begin{SIequation}
  \label{sieq:dNeom}
  \frac{\del \dNj}{\del t} =-K \nabla^{2} \dNj + S_{j},
\end{SIequation}
where $\dNj$, $j \in \{y, z\}$, are the components of the fluctuation perpendicular to the
director, and $K \equiv k/\gamma$ is the nematic diffusivity, which has units of diffusion
constant. Taking the Fourier transform of this expression, we get
\begin{equation*}
  - i \omega \dNjT = K \Qvec^{2}\dNjT+\tilde{S}_{j}(\Qvec,\omega),
\end{equation*}
where $\mathbf{Q}=\{q_{x},q_{y},q_{z}\}$ and $\omega$ are the wave-vector and angular
frequency, and the Fourier transform is defined as
\begin{equation*}
  \tilde{f}(\Qvec,\omega) \equiv FT\{ f(\Rvec,t) \} \equiv \int \dd \Qvec \, \dd \omega \,
  e^{-i(\Qvec\cdot \Rvec + \omega t)}f(\Rvec,t),
  \end{equation*}
where $f(\Rvec,t)$ is an arbitrary function of space and time.
  
\subsection{Theoretically Predicted Fourier Space Correlation Functions in (3+1)D}
To explicitly calculate correlation functions from the equation of
motion~\ref{sieq:dNeom}, we assume Gaussian-distributed noise. In real space, $S_{j}(\Rvec,t)$
is fully characterized by the equations
\begin{equation*}
  \langle S_{j}(\Rvec, t) \rangle = 0, \qquad \langle S_{i}(\Rvec, t)  S_{j}(\Rvec', t') 
\rangle = S_{0}^{2}\, \delta (\Rvec-\Rvec') \delta (t-t') \delta_{ij},
\end{equation*}
\noindent
where $S_{0}^{2}$ is the noise amplitude. These relations are equivalent to the Fourier space expressions
\begin{equation*}
  \langle \tilde{S}_{j} (\Qvec, \omega) \rangle  = 0, \qquad
  \langle \tilde{S}_{i} (\Qvec_{1}, \omega_{1}) \tilde{S}_{j} (\Qvec_{2}, \omega_{2})
  \rangle =S_{0}^{2} \, \delta(\Qvec_{1}+ \Qvec_{2}) \delta(\omega_{1}+ \omega_{2}) \delta_{ij}.
\end{equation*}
\noindent
Using these expressions, we calculate the full spatiotemporal correlation function for $\dNyT$,
\begin{SIequation}
 \label{eq:Cnem3D}
C_{\nn} (\Qvec, \omega) \equiv  \frac{1}{V_{3+1}}\dNyT(\Qvec,\omega)^{*}\dNyT(\Qvec, \omega)  = \frac{S_{0}^{2}
}{\omega^{2}  + K^{2} \Qvec^{4}}, 
\end{SIequation}
where $V_{3+1}$ is the (3+1)D system volume. The equal time correlation function is given by evaluating the inverse Fourier transform (in the time
domain only) of $C_{\nn}(\Qvec, \omega)$ at $t=0$,
\begin{SIequation}
  \label{eq:equalTimeDef3D}
S_{\nn} (\Qvec)  = \frac{1}{2\pi} \int_{-\infty}^{\infty} C_{\nn} (\Qvec, \omega)  d \omega= \frac{S_{0}^{2}}{2 K \Qvec^{2}}.
\end{SIequation}

\subsection{LC-PolScope Measurements, Projection Slice Theorem, Correlation Functions in (2+1)D}
The LC-PolScope does not allow measurement of the full 3D director
$\mathbf{\hat{N}}(\mathbf{R})$. Rather, we measure the orientation field projected over
the optical axis $\zhat = \ohat$ (Main Text Eqns.~1 \& Fig.~1(a)). According to the projection-slice theorem
of Fourier analysis, this is equivalent to taking a long wavelength limit in the $\zhat$
direction, i.e. $q_{z} \rightarrow 0$~\cite{Bracewell1990}. In operator notation,
\begin{equation*}
S_{q_{z}=0} FT_{xyz} =   FT_{xy} P_{z},
  \end{equation*}
where $S_{q_{z}=0}$ is the slice operator that sets $q_{z} \rightarrow 0$ in its argument
function, $FT_{xy}$ and $FT_{xyz}$ are 2D and 3D spatial Fourier transforms, $P_{z}$ is the
projection operator over $z$, $P_{z}\{ f(x,y,z)\} = \int f(x,y,z) \dd z $, 
and we have suppressed the time dimension for convenience. With this notation,
\begin{equation*}
\dNyT(q_{x},q_{y}, q_{z} \rightarrow 0) =  F_{xy} P_{z} \dNy \approx F_{xy}\int_{T} \dNy
\dd z \approx T \dnyT,
\end{equation*}
where $T$ is the sample thickness along $\zhat$. In the above expression, the first
approximation enters because of the finite sample thickness, and the second arises because
$\hat{n}$ is a normalized projection rather than an absolute projection, see Eqns.~1 of the
Main Text. (The last relation is accurate to first order in $\dNj$, however.) Applying
the limit $q_{z} \rightarrow 0 $ to equations \ref{eq:Cnem3D} and \ref{eq:equalTimeDef3D}
therefore yields
\begin{SIequation}
 \label{eq:corrFuncs2D}
c_{\nn} (q_{x}, q_{y}, \omega) \equiv \frac{1}{V_{2+1}} \dnyT(\qvec,\omega)^{*}\dnyT(\qvec, \omega)  = \frac{s_{0}^{2}
}{\omega^{2}  + K^{2} (q_{x}^{2} + q_{y}^{2})^{2} }; \qquad s_{\nn}(q_{x},q_{y}) = \frac{s_{0}^{2}}{2 K (q_{x}^{2}+q_{y}^{2})},
\end{SIequation}
where $V_{2+1} = V_{3+1}/T$ is the (2+1)D system volume, and $s_{0}^{2} \equiv
S_{0}^{2}/T$. As shown in~\cite{Brugues2014}, In the long wavelength limit in the
direction along the spindle axis, $q_{x}\rightarrow 0$, $s_{\nn}$ takes a particularly
simple inverse-square power law form,
\begin{SIequation}
  \label{eq:snnInfiniteDef}
s_{\nn}(0,q_{y}) = \frac{s_{0}^{2}}{2 K q_{y}^{2}}.
\end{SIequation}

\subsection{Comparison of Experimental Correlation Functions
  with Theoretical Predictions; Estimate of Position of Peak in $s_{\nn}(q_{y})$}
\begin{figure}[t!]
  \centering
  \includegraphics[width=\textwidth]{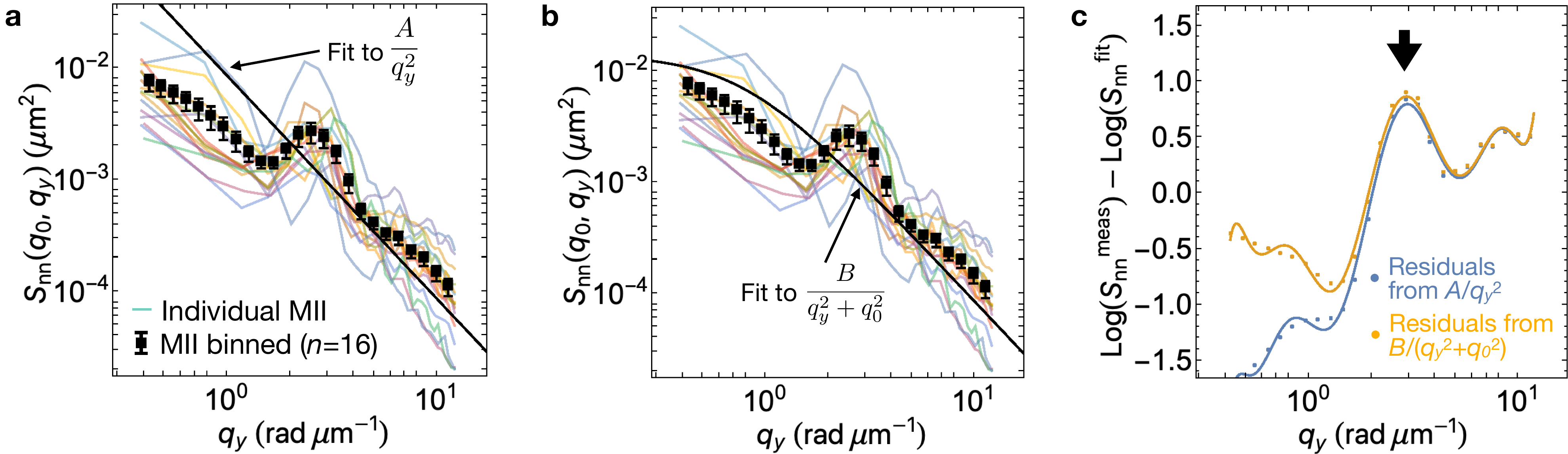}
  \caption{ \label{fig:snnFits} Fits of theoretical predictions of $s_{\nn}$ to $s_{\nn}$
calculated from LC-PolScope fluctuation data. (a) Fit of avergage experimentally
determined $s_{\nn}(q_{0}, q_{y})$ (black points with error bars) to theoretical
prediction, assuming $q_{0} \rightarrow 0$ (Eqn.~\ref{eq:snnInfiniteDef}). (b) Fit of
avergage experimentally determined $S_{\nn}(q_{0}, q_{y})$ to theorietical prediction,
assuming $q_{0} = 2\pi/\lambda_{0}$ (Eqn.~\ref{eq:corrFuncs2D}), where $\lambda_{0} = 8
\um$ is the size of the analysis box. (See Main Text Fig.~2.) (c) The location of the peak
(black arrow at $q_{y}= 2.9$ rad$\um^{-1}$) is identified by fitting the residuals (blue
and yellow points) of the fits in (a) and (b), calculated in log space. Continuous curves
are ad-hoc polynomial fits; the peak near $q_{y}= 3.0$ rad$\um^{-1}$ is identified using
Newton's method.}
\end{figure}

Fits of Eqns.~\ref{eq:corrFuncs2D}~and~\ref{eq:snnInfiniteDef} to the correlation
functions calculated from LC-PolScope data are shown in S.I.~Fig.~\ref{fig:snnFits}(a \& b). We
use these fits to estimate the location $q_{y}^{*}$ of the anomalous peak in
$s_{\nn}(q_{0},q_{y})$ as follows: we first calculate, in log space, the residuals
corresponding to each fit (S.I.~Fig.~\ref{fig:snnFits}(c)). We then use an ad-hoc fit (a
polynomial of degree 10) to obtain a continuous approximation of the residual as a
function of $q_{y}$, and use Newton's method to identify the value of $q_{y}$
corresponding to its maximum value. The value of $q_{y}^{*}$ we obtain using this method
is not sensitive to paramaters such as the number of bins used to calculate the average
experimental $s_{\nn}$ or the specific form of the ad-hoc approximation to the residual
curve. To obtain the uncertainty in $q_{y}^{*}$ reported in the Main Text, we perform
empirical bootstrapping over single-spindle data sets: we first randomly select 50\% of
the single-spindle $s_{\nn}$ curves (i.e.~50\% of the faint colored curves in
S.I.~Fig.~\ref{fig:snnFits}(a)) and obtain a value of $q_{y}^{*}$ using only that data. We then
repeat this procedure 100 times, and take the standard deviation of the resulting
distribution of $q_{y}^{*}$ values to be the uncertainty in $q_{y}^{*}$.

\subsection{Regularly Spaced Voids in a $q^{-2}$-Correlated Background Generate a Peak in
  $s_{\nn}(q)$: Simulations}
\begin{figure}[t!]
  \centering
  \includegraphics[width=\textwidth]{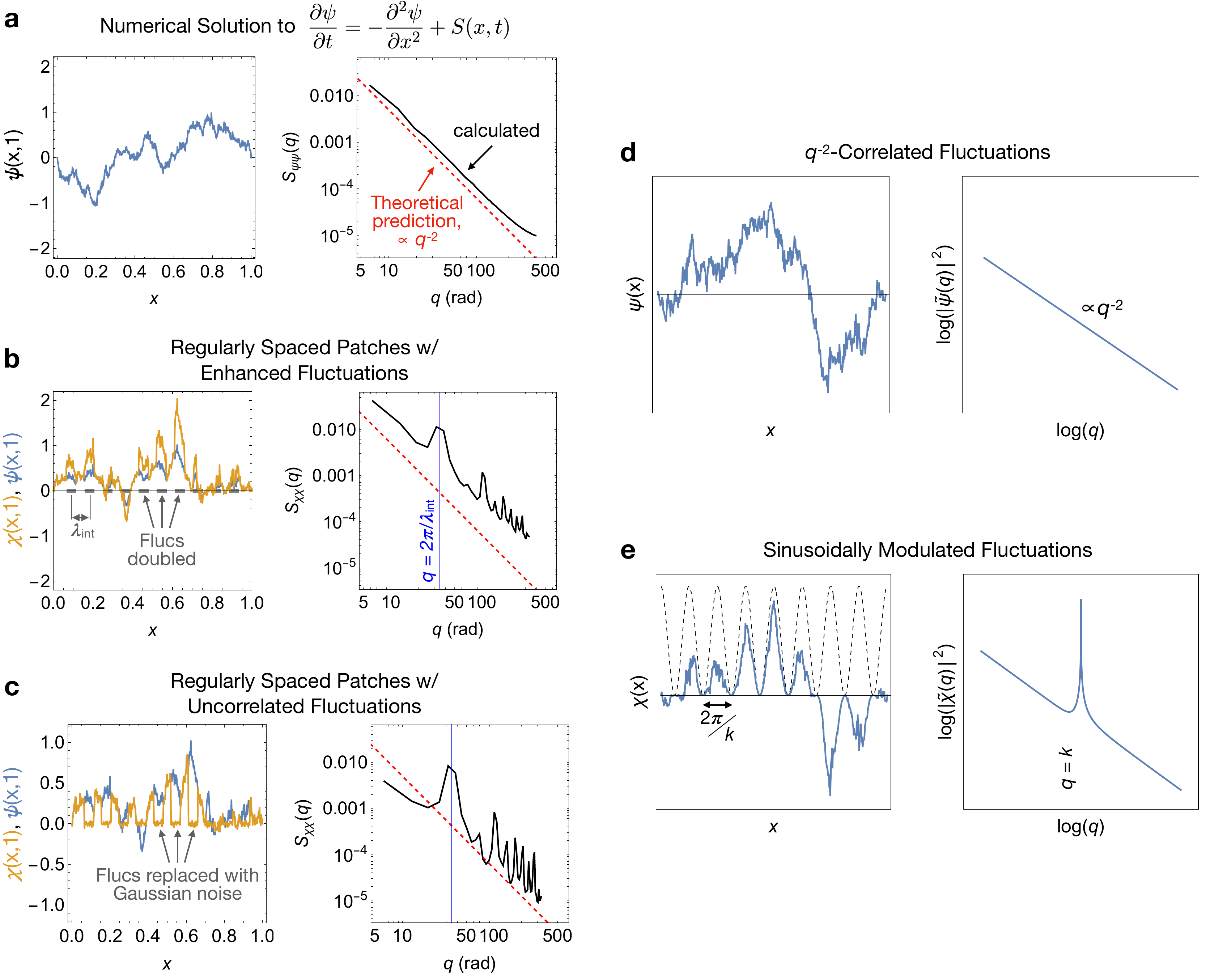}
  \caption{ \label{fig:flucsWvoids} Regularly spaced patches in a fluctuating background
cause a peak in the equal-time correlation function at the wavenumber corresponding to the
spacing between voids. (a) \textit{Left:} Numerical solution $\psi(x,t)$ of the (1+1)-D
diffusion equation with non-conservative noise, plotted at a time $t=1$ that is long
enough for the simulation to sample the steady-state distribution of field
configurations. \textit{Right:} The calculated equal-time correlation function $S_{\psi
\psi}(q)$ (black curve) closely matches the theoretical prediction, $S_{\psi \psi}(q) =
1/(2 q^{2})$ (red dashed line). (b) \textit{Left:} Starting with the numerical solution
$\psi(x,t)$, we define a new field $\chi(x,t)$ that has larger fluctuations in $N = 10$
regularly spaced patches, each of width $w_{0}=0.05$ (grey bars along
$x$-axis). \textit{Right:} The corresponding equal-time correlation function has a peak at
the $q$-value corresponding to the period $\lambda_{\text{int}}$ of the patches. The
overall amplitude $S_{\chi \chi}(q)$ of the curve is larger due to the larger
fluctuations, but retains its $q^{-2}$ shape at high and low $q$. The additional spikes at
higher $q$ are due to ``ringing'', a numerical artifact in which the sharp periodic jumps
in $\chi(x,t)$ cause peaks at integer multiples of $2\pi/\lambda_{\text{int}}$ when the
Fourier transform is calculated. (c) \textit{Left:} Starting with the numerical solution
$\psi(x,t)$, we define a new field $\chi(x,t)$ that has Gaussian (i.e.~uncorrelated)
fluctuations in 10 regularly spaced patches. \textit{Right:} The corresponding equal-time
correlation function has a peak at the $q$-value corresponding to the period of the
patches. (d) A field $\psi(x)$ has the power spectrum $|\tilde{\psi}(q)|^{2} \propto
q^{-1}$. (e) Modulating $\psi(x)$ by a sinusoidal function of period $2\pi/k$ leads to a
peak in the power spectrum of the modulated function at $q=k$, and $q^{-2}$ behavior at
higher and lower $q$.}
\end{figure}

To understand in greater detail how voids in the microtubule network lead to the observed
peak in $s_{\nn}(q_{0},q_{y})$ (Main Text Fig.~2(c)), we performed a series of 
simulations (S.I.~Fig.~\ref{fig:flucsWvoids}(a-c)). To model fluctuations in the
microtubule network in the absence of voids, we used a forward time-centered space (FTCS)
scheme to numerically integrate a dimensionless (1+1)-D version of Eqn.~\ref{sieq:dNeom}
for fluctuations in a contiguous active nematic, i.e.~the 1D
diffusion equation with non-conservative Gaussian noise,
\begin{SIequation}
  \label{sieq:diff1D}
\frac{\del \psi}{\del t} =-\frac{\del^{2} \psi}{\del x^{2}}  + S(x,t),
\end{SIequation}
where $\psi(x,t)$ is the fluctuating field and the scalar noise $S(x,t)$ is defined by the
equations $\langle S(x, t) \rangle = 0$, $\langle S(\xvec, t) S(\xvec', t') \rangle =
\delta (x-x') \delta (t-t')$. Choosing a time-step that satisfies the stability criterion
$\Delta t \leq \Delta x^{2}/2$, we integrate this equation on the spatio-temporal domain
$[0,1] \times [0,2]$; this choice of domain allows relaxation of the longest-wavelength modes. We
impose Dirichlet boundary conditions $\psi(0,t) = \psi(1,t) = 0$ for all $t$, and
initialize the simulation with $\psi(x,0) =0$ for all $x$. In all subsequent analysis, we
consider only the values of $\psi(x,t)$ for $t \in [1,2]$, for which the
simulation samples the steady-state distribution of field configurations. A snapshot of
the fluctuating field at $t=1$ is shown in S.I.~Fig.~\ref{fig:flucsWvoids}(a), as is the
equal-time correlation function $S_{\psi \psi}(x)$, averaged over all $t \in [1,2]$; the
latter plot decays as $q^{-2}$ as predicted by Eqn.~\ref{eq:snnInfiniteDef}.

To model the effect of enhanced fluctuations in regularly spaced voids, we construct a new
field $\chi(x,t)$ by replacing the simulated values of $\psi(x,t)$ in $N$ periodically
spaced intervals ("patches"), whose centers $c_{i} = \{ i/(N+1) \}_{i=1,N} $ are separated by
wavelength $\lambda_{\text{int}} = 1/(N+1)$, and which have width $w_{0}
<\lambda_{\text{int}}$. In those intervals, $\chi(x,t)$ is doubled,
 \[ \chi(x,t) = \begin{cases} 
         2\psi(x,t) & x \in (c_{i}-\frac{w_{0}}{2}, c_{i}+\frac{w_{0}}{2}) \text{ for any } i, \\
          \psi(x,t) & \text{otherwise.}
       \end{cases}
    \]
Having constructed a real-space model of a field with increased fluctuations in regularly
spaced patches, we next compute its equal-time correlation function $S_{\chi \chi}(q)$
(S.I.~Fig.~\ref{fig:flucsWvoids}(b)). For appropriate choice of the parameters $N$ and
$w_{0}$, $S_{\chi \chi}(q)$ displays a peak at $2 \pi/\lambda_{\text{int}}$ superimposed
on a $q^{-2}$ decay that predominates at both higher and lower $q$-values, similar to the
form of the $S_{\nn}(q_{0}, q_{y})$ plot calculated from experimental observations (Main
Text Fig.~2(c)). A similar result can be recovered if, instead of doubling the field in
the patches, the correlated fluctuations are replaced by spatially
uncorrelated Gaussian white noise; this implies that the peak in $S_{\chi \chi}$ is not
sensitive to the specific behavior of $\chi(x,t)$ in the patches
(S.I.~Fig.~\ref{fig:flucsWvoids}(c)).

\subsection{Regularly Spaced Voids in a $q^{-2}$-Correlated Background Generate a Peak in
  $s_{\nn}(q)$: Analytical Calculation}
By using the convolution theorem of Fourier analysis, the emergence of peaks in
$S_\nn(q_{0},q_{y})$ can also be understood from a theoretical perspective. Let $\psi(x)$
be a real-valued function with the power spectrum $|\tilde{\psi}(q)|^{2} =
|A_{0}|^{2}q^{-2}$. (For functions of both space and time, the time-averaged spatial power spectrum is
identical to the equal-time correlation function $S_{\psi \psi}(q)$ apart from a constant
normalization factor.) This form of $|\tilde{\psi}(q)|^{2}$ implies that $\tilde{\psi}(q) =
A_{0} q^{-1}$, where $A_{0}$ may be complex-valued. We represent periodic voids by
multiplying $\psi(x)$ by a sinusoidal function with minimum zero and maximum 1,
\begin{equation*}
\chi(x) \equiv  \psi(x) (1 - \sin{k x})/2,
\end{equation*}
where $k$ is the angular wavenumber associated with the spatial period
of the sinusoid. Taking the Fourier transform of $\chi(x)$ and applying the
convolution theorem yields
\begin{equation*}
\tilde{\chi}(q) =  \frac{1}{2}\Big{(} \tilde{\psi}(q) - FT\{\psi(x) \sin{k x} \} \Big{)} =
\frac{1}{2}\Big{(} \tilde{\psi}(q) - \frac{1}{2 i}(\tilde{\psi}(q-k) - \tilde{\psi}(q+k)) \Big{)}.
\end{equation*}
From this expression we can directly calculate the power spectrum $|\tilde{\chi}(q)|^{2}$,
\begin{equation*}
|\tilde{\chi}(q)|^{2} = \frac{1}{4} \Big{(} |\tilde{\psi}(q)|^{2} +
\frac{1}{4}(|\tilde{\psi}(q-k)|^{2} + 2|\tilde{\psi}(q-k)||\tilde{\psi}(q+k)| +
|\tilde{\psi}(q+k)|^{2})  \Big{)} = \frac{|A_{0}|^{2}}{4}\frac{q^{4} + (q^{2}-k^{2})^{2}}{q^{2}(q-k)^{2}(q+k)^2},
\end{equation*}
which has a divergence at $q=k$ and is proportional to $q^{-2}$ for both $q \gg k$ and $q
\ll k$, as expected (S.I.~Fig.~\ref{fig:flucsWvoids}(d-e)). 

\newpage

\section{Correlations Between Retardance and Fluctuation Magnitude, Corrected for Varying Spindle Thickness}
\begin{figure}[t!]
  \centering
  \includegraphics[width=\textwidth]{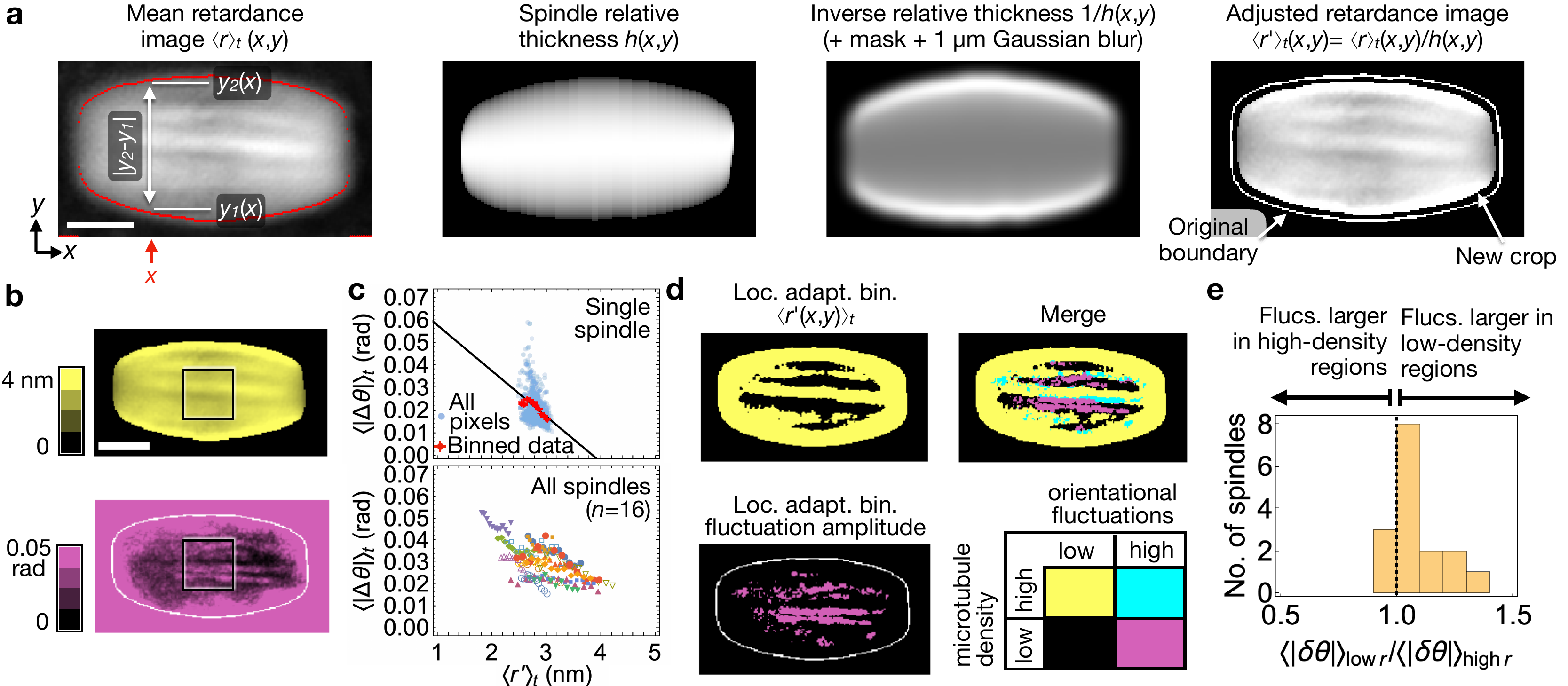}
  \caption{ \label{fig:adjRetFig} Correlations between fluctuation magnitude and adjusted
retardance.  (a) Finding the time-averaged retardance image $\langle r' \rangle_{t}(x,y)$
adjusted to take account of varying spindle thickness. From left to right, we first
identify the spindle midpoint $(y_{1}+y_{2})/2$ and radius $|y_{2}-y_{1}|/2$ at each point
$x$ along the spindle axis. From these values, we construct the normalized relative
thickness image $h(x,y)$, which takes the value 1 at the midpoint between the vertical
positions of the spindle boundary. We next find the inverse relative thickness image
$1/h(x,y)$; to avoid pixelization artifacts, we first multiply the raw $1/h(x,y)$ image by
the retardance mask (the region inside the boundary shown in the $\langle
r\rangle_{t}(x,y)$ image), and convolve the resulting image with a Gaussian filter of
radius $1 \um$. We find the adjusted retardance image $\langle r' \rangle_{t}(x,y)$ by
pixelwise multiplication and finally multiply this image by a new retardance mask
constructed by removing $1 \um$ from all boundaries of the original mask. (b-e) Once we
have constructed $\langle r' \rangle_{t}(x,y)$, we repeat without further modification the
analysis shown in Main Text Fig.~2.}
\end{figure}

In Fig.~2 of the Main Text and associated discussion, we demonstrated that the
time-averaged retardance $\langle r \rangle_{t}$ at each pixel is negatively correlated
with the time-averaged magnitude of orientational fluctuations $\langle |\delta \theta|
\rangle_{t}$. However, without further analysis, it is possible that this is an artifact
caused by the fact that retardance is proportional to sample thickness: For an
approximately axisymmetric shape like the spindle, with its rotational axis in the image
plane, the sample thickness will be largest at the axis and decay to zero at the
boundaries. Thus, if the spindle were uniform density, observed retardance would be
highest along the long axis of the spindle $y=0$ and lowest at the spindle boundary
(Eqn.~\ref{sieq:retProfile}; Main Text Fig.~3(d)). At the same time, because of
interactions between microtubules and the spindle boundary, orientational fluctuations
could in principle be either enhanced or suppressed near the edge of the spindle, which
might lead to (positive or negative) correlations between retardance and orientaiontal
fluctuations that are not due to a direct relationship between them, but instead arise
from the fact that that both depend on distance from the boundary. To remove this
potential artifact, we constructed the corrected retardance
\begin{equation*}
\langle r' \rangle_{t}(x,y) \equiv  \frac{\langle r\rangle_{t}(x,y)}{h(x,y)},
\end{equation*}
where $h(x,y)$ is the relative $z$-thickness at position $(x,y)$ in the spindle
(S.I.~Fig.~\ref{fig:adjRetFig}(a)). To estimate $h(x,y)$, we first identify the
$y$-coordinates $y_{1}(x)$ and $y_{2}(x)$ of the spindle boundary at a given position $x$
along the spindle long axis. For that value of $x$, $h(x,y)$ is then given by the equation
of the arc of an ellipse with maximum height unity and values of zero at each boundary,
\begin{equation*}
h(x,y) =\sqrt{ 1 -  \Big( \frac{-2 y + y_{1}+y_{2}}{y_{1}-y_{2}} \Big)^{2}}.
\end{equation*}
To avoid the singular values of $\langle r'\rangle_{t}(x,y)$ that appear (by construction)
at the spindle boundary, we construct a tighter mask by removing $1 \um$ from the original
mask. After making these modifications, repeating the analysis shown in Fig.~2 of the Main
Text with $\langle r'\rangle_{t}(x,y)$ instead of $\langle r\rangle_{t}(x,y)$ yields
results that are identical within experimental uncertainty (S.I.~Fig.~\ref{fig:adjRetFig}(b-e); S.I.~Table~\ref{tab:rVsr1vals}).

\begin{table}[h!]
\centering
\begin{tabular}{ |c||c|c| } 
 \hline
  & Analysis using $\langle r\rangle_{t}(x,y)$ & Analysis using $\langle r'\rangle_{t}(x,y)$ \\
   \hline  \hline
 \makecell{fraction of oocytes with \\ negative slope of $\langle |\delta \theta|
  \rangle_{t}$ \\ vs (adjusted) retardance }  & 15/16 & 15/16 \\
   \hline
\makecell{average slope of $\langle |\delta \theta|\rangle_{t}$ \\ vs (adjusted)
  retardance \\ (mean $\pm$ SE,  rad/nm$^{-1}$ )}   & $-0.009 \pm 0.003$ & $-0.010\pm 0.002$ \\
   \hline
  \makecell{average value of \\
  $\langle |\delta \theta|\rangle_{\text{low } r}/$ $\langle |\delta
  \theta|\rangle_{\text{high } r}$\\  (mean $\pm$ SE)}& $1.07 \pm 0.02$ & $1.08 \pm 0.03$ \\
 \hline
\end{tabular}
\caption{Orientational fluctuation vs retardance: comparison of parameter values obtained
using time-averaged retardance $\langle r\rangle_{t}(x,y)$ vs time-averaged adjusted
retardance $\langle r'\rangle_{t}(x,y)$.}
\label{tab:rVsr1vals}
\end{table}

\newpage

\section{Analysis of Confocal Micrographs of Metaphase Plate}
\subsection{Finding Chromosome Shape Parameters}
\label{sisec:chrShapes}
\begin{figure}[t!]
  \centering
  \includegraphics[width=\textwidth]{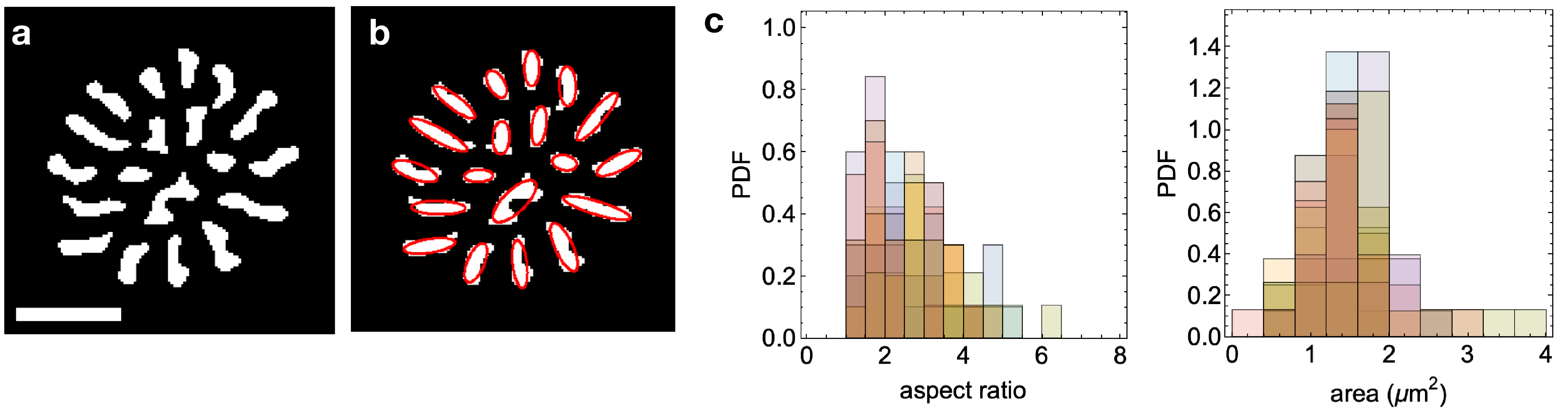}
  \caption{ \label{fig:ellipseFits} Properties of chromosome sections in the metaphase
plate. (a \& b) Best fit ellipses overlaid on binarized chromosome sections for one
spindle. (c) Distribution of chromosome aspect ratios $a/b$ and areas in $n=11$ spindles,
in each of which we have identified $\nChr = $19 or 20 chromosome sections. The mean value
of $a/b$ is $2.5 \pm 0.1$, and the average chromosome area is $(1.45 \pm 0.03) \um^{2}$
(mean $\pm$ SE).}
\end{figure}

To estimate the shapes of chromosome sections in the metaphase plate ($x=0$ plane), we
begin with the binarized sections shown in Fig.~4(a) of the Main Text. Let
$\mathscr{C}_{i}$ be the geometrical region corresponding to chromosome section $i$;
define $\eIi$ and $\eIIi$ to be the eigenvectors of the moment of inertia tensor of
$\mathscr{C}_{i}$ about its center of mass, with corresponding eigenvalues $e_{i,1}$ and
$e_{i,2}$, where $e_{i,1} > e_{i,2}$. The orientation of the best fit ellipse is given by
$\eIi$ while its major and minor axes are given by
\begin{equation*}
a_{i} = 2\Big(\frac{e_{i,1}^{3}}{e_{i,2}} \Big)^{1/8}; \qquad b_{i} = \frac{2 e_{i,1}}{(a/2)^{3}}.
\end{equation*}
The results of this procedure, applied to a set of binarized chromosomes sections, are
shown in S.I.~Fig.~\ref{fig:ellipseFits}, which also gives histograms of chromosome shape
parameters for $n=11$ MII spindles. For a given spindle, the average 
shape parameters are defined as follows:
\begin{SIequation}
  \label{sieq:chrShapes}
  a \equiv \langle a_{i} \rangle_{i=1...\nChr}; \qquad b\equiv  \langle b_{i} \rangle_{i=1...\nChr},
\end{SIequation}
where $\nChr$ is the number of identified chromosome sections in that spindle.

\subsection{Definition of $g_{II}(s)$}
\begin{figure}[t!]
  \centering
  \includegraphics[width=0.5\textwidth]{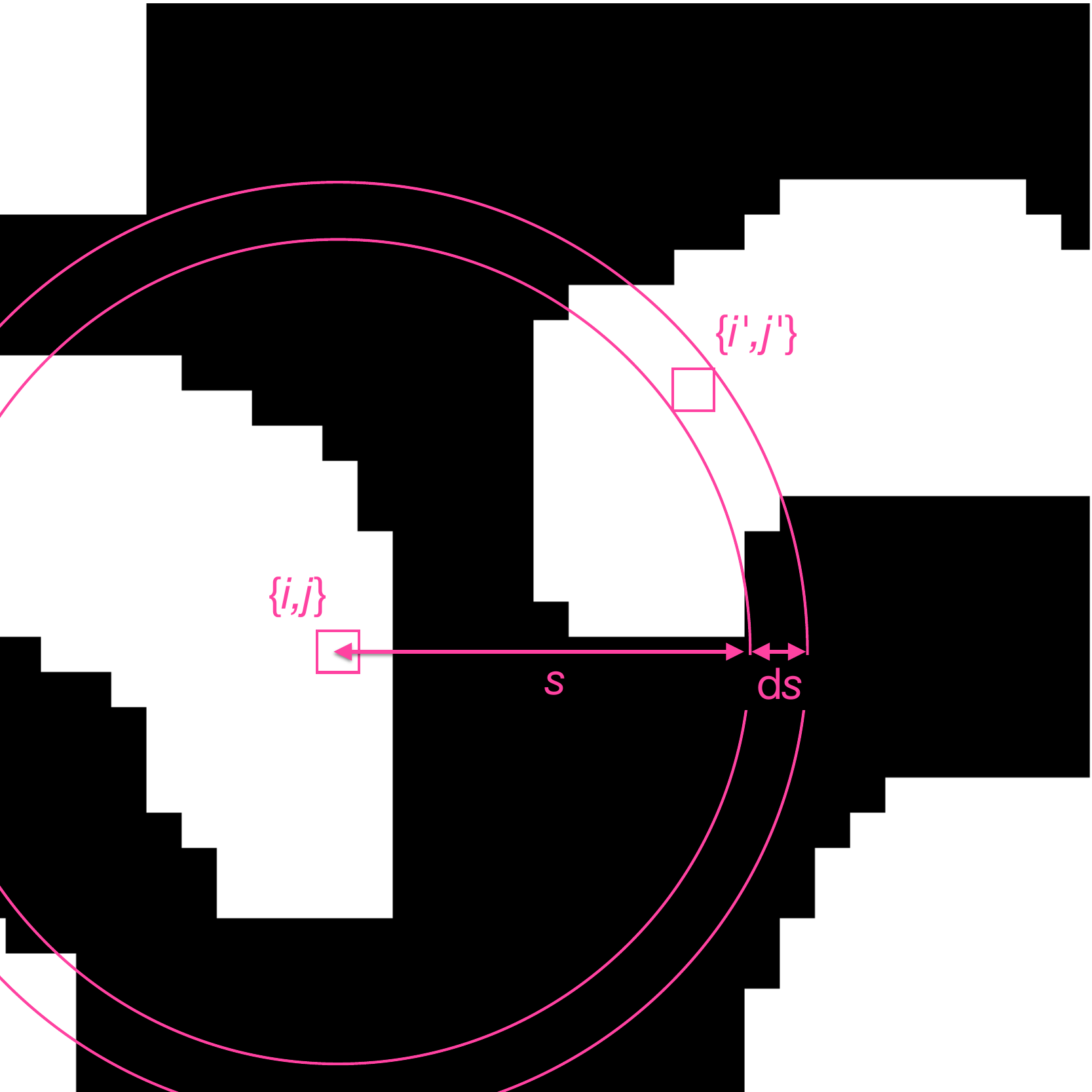}
  \caption{\label{fig:gIIofs} Calculation of $g_{II}(s)$. We first draw an annulus
centered at $\{i,j\}$, with inner radius $s$ and outer radius $s + ds$, and then calculate
$\psi_{s}(i,j)$, the fraction of white pixels in the annulus. Then, $g_{II}(s) $ is the
average of $\psi_{s}(i,j)$ over all white pixels, i.e.~all $\{i,j\}$ with $I(i,j) = 1$,
normalized by the fraction of white pixels over the whole image. }
\end{figure}

To quantify spatial correlations in pixel intensity in binarized images, we use the
intensity-intensity correlation function $g_{II}(s)$, an adaptation of the pair
correlation function commonly used in condensed matter physics~\cite{Satoh2003}. Let
$I(i,j)$ be the value of the pixel at position $\{i,j\}$ in an image; in a binarized
image, $I(i,j) = 0$ or 1 (S.I.~Fig.~\ref{fig:gIIofs}). To calculate $g_{II}(s)$, we consider
all pixels $I(i',j')$ where $\{i',j'\}$ lies in an annulus, centered at $\{i,j\}$, of inner
radius $s$ and outer radius $s + ds$, where $ds$ is of order one pixel. Let
$\psi_{s}(i,j)$ be the fraction of white pixels in the annulus, $\psi_{s}(i,j) = \langle
I(i',j') \rangle_{s<||\{i,j\}-\{i',j'\}||<s+ds}$. Then, $g_{II}(s) $ is the average of
$\psi_{s}(i,j)$ over all white pixels, i.e.~all $\{i,j\}$ with $I(i,j) = 1$, normalized by
the fraction of white pixels over the whole image.
\newpage

\section{Simulations of Chromosome Configurations}
\subsection{Generating Randomized Non-Overlapping Chromosome Configurations}
\begin{figure}[t!]
  \centering
  \includegraphics[width=\textwidth]{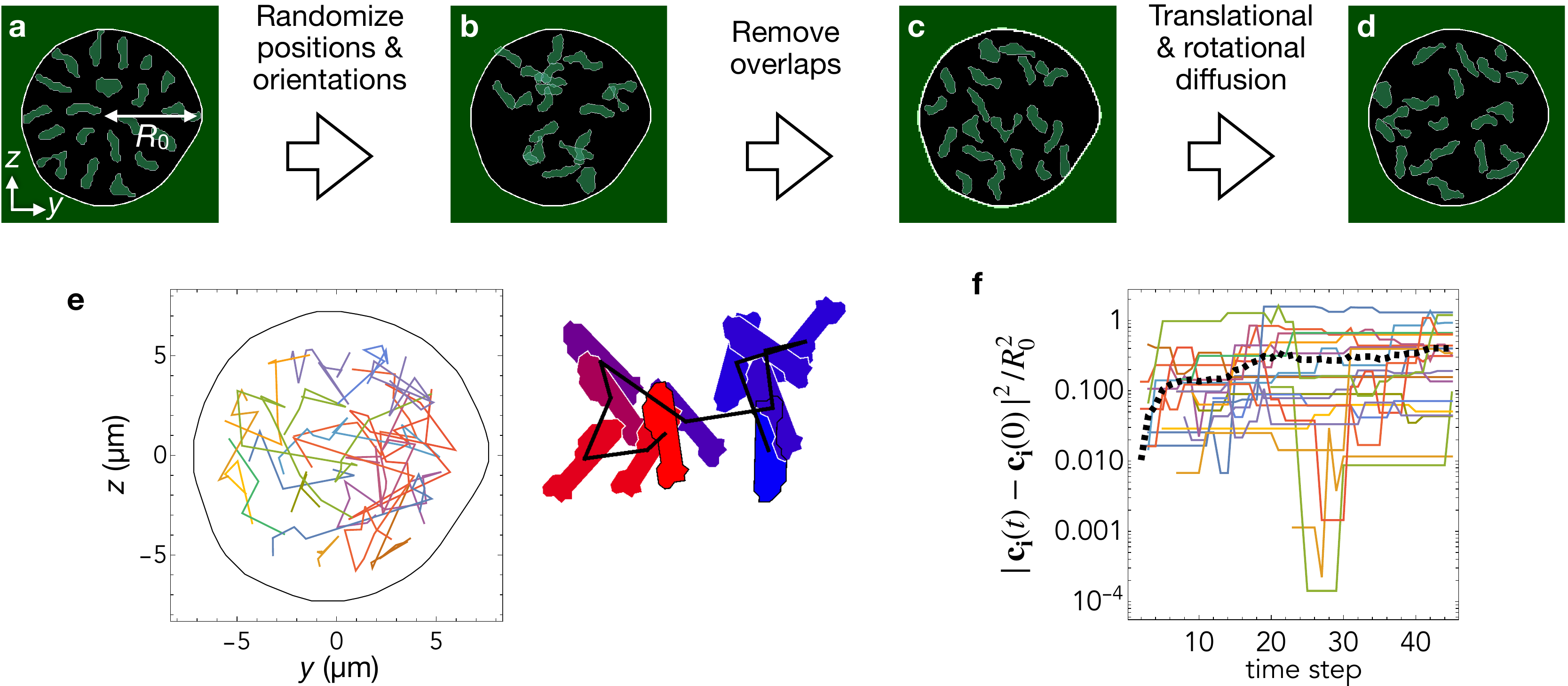}
  \caption{\label{fig:monteCarloSims} Overview of simulations to generate randomized
non-overlapping chromosome configurations. (a-b) We identify the outlines of $\nChr$
chromosome sections and randomly place them inside the spindle boundary; (b-c) we
translate and rotate chromosomes sections to reduce and ultimately eliminate the overlap
of chromosome sections with each other and with the region outside the spindle boundary;
(c-d) we randomly translate and rotate non-overlapping chromosome sections to generate free
diffusion. (e) \textit{Left:} Trajectories of all centroids $\mathbf{c_{i}}(t)$ during the
free diffusion stage of the simulation. Each color corresponds to the trajectory of the
centroid of one chromosome section. \textit{Right:} Trajectory of a single
chromosome section $\mathscr{C}_{i}$ during the free diffusion stage. Blue indicates the
initial configuration; red indicates the final configuration; the black line indicates the
trajectory of the centroid $\mathbf{c_{i}}(t)$. (f) Square displacement of the centroids,
in units of the metaphase plate radius, defined here as $R_{0} \equiv \sqrt{A_{MP}/\pi}$,
where $A_{MP}$ is the area inside the spindle boundary. The free diffusion stage of the
simulation proceeds until the centroids of all sections have translated at least $0.1
R_{0}$ and the average absolute displacement is at least $0.5 R_{0}$. (Black dashed line
in plot shows the average square displacement from initial position.)}
\end{figure}

To compare the observed chromosome configuration with the configuration we would expect if
non-interpenetrating chromosomes were randomly positioned in the metaphase plate, we
performed a series of Monte Carlo simulations to generate randomized chromosome
configurations. The code takes as input a set of $\nChr$ binarized chromosome
sections, $\{\mathscr{C}_{i}\}_{i=1...\nChr}$, where $\mathscr{C}_{i}$ is the
geometrical region corresponding to chromosome section $i$, as well as a region
$\mathscr{M}$ corresponding to the metaphase plate (Main Text Fig.~4,
S.I.~Fig.~\ref{fig:monteCarloSims}). In the \textit{Mathematica} programming language,
the {\tt Region[]} function provides a convenient way to store and perform computations on
the geometrical regions generated by image binarization. Let $\mathbf{c}_{i}$ and
$\phi_{i}$ be the center of mass and orientation of the region $\mathscr{C}_{i}$. In pseudocode,
our simulation proceeds: \newline

\noindent Stage 1: Generate a Randomly Seeded Non-Overlapping Configuration of Chromosomes
  (S.I. Fig.~\ref{fig:monteCarloSims}(a-c))
\begin{enumerate}
  \item Randomly place chromosome sections in the plane such that
    the center of mass of region $\mathscr{C}_{i}$ is $\mathbf{c}_{i}$ and its orientation is
    $\phi_{i}$, where $\phi_{i}$ is defined relative to an arbitrary
    but fixed axis in $\mathscr{C}_{i}$. Centers and orientations
    are random variables defined such that
  \begin{equation*}
    \mathbf{c}_{i} \text{ is uniformly distributed in } \mathscr{M}; \qquad \phi_{i} \text{ is uniformly distributed in } [0,2\pi).
  \end{equation*}
  
  \item Define the total area of intersection of chromosome $i$ with all
other chromosomes and with the spindle exterior,
  \begin{equation*}
   A_{\text{int}}(i) = \sum_{ j \neq i}A(\mathscr{C}_{i} \cap \mathscr{C}_{j}) + A(\mathscr{C}_{i} \cap \mathscr{M}^{C}),
    \end{equation*}
    where $\mathscr{M}^{C}$ is the complement of $\mathscr{M}$ and
    $A(\mathscr{R})$ is the area of region $\mathscr{R}$.
    
\item If $A_{\text{int}}(\mathscr{C}_{1}) = 0$, proceed to the next step. Otherwise,
define a new region, $\mathscr{C}_{1}'$, by rotating $\mathscr{C}_{1}$ by an angle $\Delta
\phi$ about $\mathbf{c}_{1}$ and translating the resulting region by a vector $\Delta
\mathbf{c}$, where $\Delta \phi$ and each component of $\Delta \mathbf{c}$ and are
randomly drawn from Gaussian distributions with mean zero and amplitude $\Delta \phi_{0}$
and $\Delta c_{0}$ respectively. To minimize the run-time of the simulation, we set
$\Delta c_{0} = 2 \um $ and $\Delta \phi_{0} = \pi/4$, but different choices for these
parameters does not affect the statistics of the final configuration. If
$A_{\text{int}}(\mathscr{C}_{1}') < A_{\text{int}}(\mathscr{C}_{1})$, then set
$\mathscr{C}_{1} \rightarrow \mathscr{C}_{1}'$. Otherwise, proceed to the next step.

\item Repeat the previous step for chromosomes
  2, ... $\nChr$. Then return to chromosome 1, and iterate
  until $A_{\text{int}}(\mathscr{C}_{i})=0$ for all $i$ (S.I.~Fig.~\ref{fig:monteCarloSims}(c))).
  
\end{enumerate}

  \noindent{Stage 2: Monte Carlo Simulation: Generate Translational and Rotational Diffusion of Non-Overlapping
    Chromosomes (S.I.~Fig.~\ref{fig:monteCarloSims}(c-d))}
  
  \begin{enumerate} \setcounter{enumi}{4}

    \item Let $\mathbf{c}_{i}^{\text{ini}}$ and $\phi_{i}^{\text{ini}}$ be
      the center of mass and orientation of region $\mathscr{C}_{i}$
      at the completion of Stage 1.

  \item Define a region, $\mathscr{C}_{1}'$, by rotating and translating $\mathscr{C}_{1}$
by $\Delta \phi$ and $\Delta \mathbf{c}$, as defined in Step 3. If
$A_{\text{int}}(\mathscr{C}_{1}') =0 $, then set $\mathscr{C}_{1} \rightarrow
\mathscr{C}_{1}'$. Otherwise, proceed to the next step.

\item Repeat the previous step for chromosomes 2, ... $\nChr$.

  \item Let $\mathbf{c}_{i}^{\text{new}}$ and $\phi_{i}^{\text{new}}$ be the center of the
center of mass and orientation of chromosome $\mathscr{C}_{i}$. If the following three
conditions are met: (i) For all $i$, $||
\mathbf{c}_{i}^{\text{new}}-\mathbf{c}_{i}^{\text{ini}}|| > 0.1 R_{0} $; (ii) $\langle ||
\mathbf{c}_{i}^{\text{new}}-\mathbf{c}_{i}^{\text{ini}}|| \rangle > 0.5 R_{0} $; (iii) $\langle
|| \phi_{i}^{\text{new}}-\phi_{i}^{\text{ini}}|| \rangle > \pi/4$, then set
$\mathbf{c}_{i}^{\text{fin}} = \mathbf{c}_{i}^{\text{new}}$ and $\phi_{i}^{\text{fin}} =
\phi_{i}^{\text{new}}$ and terminate the simulation. Otherwise, return to step 6.
  \end{enumerate}

Typical chromosome trajectories during Stage 2, plus the corresponding time course of the square
displacement, are shown in S.I.~Fig.~\ref{fig:monteCarloSims}(e \& f).
 
\subsection{Repulsive Ellipses: Simulation Framework for Arbitrary Potential $U(s)$}
\label{sec:repulsiveEllipses}
For director-aligned 3D voids with circular $x=0$ cross-sections, we could model the
pairwise interaction between them via a pairwise central potential $U(s)$, where $s$ is the
distance between void centers in the $x=0$ plane. \footnote{Even for
  axisymmetric, tactoid-shaped voids, however, there is no guarantee that such a model would
  capture the physics of void self-organization: in general, forces
  mediated by 3D nematic elasticity are not pairwise additive.} In that case, for pairs of voids with
centers located at $yz$-coordinates $\mathbf{s_{i}}$ and $\mathbf{s_{j}}$, the force from
void $i$ on void $j$ would be given by
\begin{equation*} \mathbf{F}_{ij} = F(s_{ij})\shat_{ij}= - \frac{\partial U}{\partial
s_{ij}}\shat_{ij},
\end{equation*} where $s_{ij} \equiv |\mathbf{s}_{j} - \mathbf{s}_{i}|$ and $\shat_{ij}
\equiv (\mathbf{s}_{j} - \mathbf{s}_{i})/|\mathbf{s}_{j} - \mathbf{s}_{i}|$.

\begin{figure}[t!]
  \centering
  \includegraphics[width=0.6\textwidth]{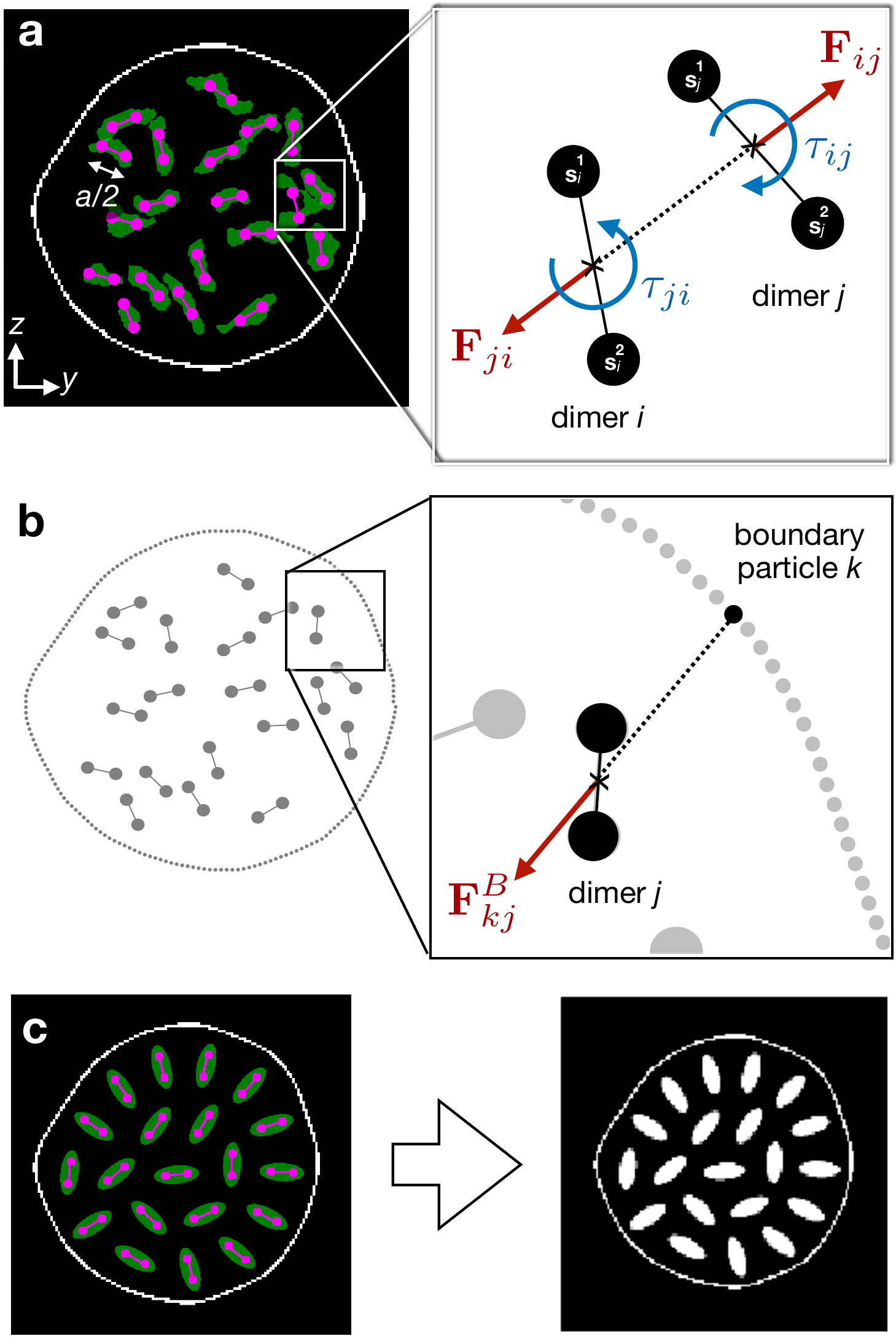}
  \caption{\label{fig:repulsiveSims} Framework for repulsive dimer simulations. (a)
Chromosome section $j$ is represented as a dimer, a pair of points $\mathbf{s}_{j}^{1}$
and $\mathbf{s}_{j}^{2}$ separated by a fixed distance $a/2$, where $a$ is the average
chromosome long axis in the $x=0$ plane. Forces and torques on dimer $j$ from dimer $i$
are inherited from the forces between the points comprising the dimers. (b) The boundary
of the spindle is represented by a chain of $N_{\text{bound}}$ equally-spaced points; each
boundary point $k$ exerts a force $\mathbf{F}_{kj}^{B}$ on the centroid of dimer $j$; the
total force $\mathbf{F}_{\text{bound},j}$ from the boundary on dimer $j$ is given by
$\sum_{k=1}^{N_{\text{bound}}}F_{kj}^{B}$. (c) To compare simulated configurations with
experimental ones at the end of the simulation, dimers are converted into identical
ellipses with major and minor axes $a$ and $b$.}
\end{figure}

For voids with non-circular $x=0$ cross-sections, long-range repulsive interactions
generically produce torques as well as forces between voids, which predicts alignment
between neighboring chromosome sections. This prediction is consistent with the
co-alignment of neighboring chromosomes evident in micrographs of the metaphase plate
(e.g.~Main Text Figs.~3(b) \& 4(a)). To model this effect, we represent the $x=0$ void cross-sections
as dimers, pairs of rigidly connected points separated by the length of the average
chromosome semi-major axis $a/2$ (S.I.~\ref{sisec:chrShapes}). Forces and torques
arise from the interactions between the points comprising one dimer with points comprising
another (S.I.~Fig.~\ref{fig:repulsiveSims}(a)). More specifically, void section $i$ is
represented by two points $\mathbf{s}^{1}_{i}$ and $\mathbf{s}^{2}_{i}$, separated by the
constant length $a/2$, i.e.~$|\mathbf{s}^{2}_{i}-\mathbf{s}^{1}_{i}| = a/2$. The center of
mass of void section $i$ is $\mathbf{s}^{\text{CoM}} =
(\mathbf{s}^{1}_{i}+\mathbf{s}^{2}_{i})/2$ and its orientation $\phi_{i}$ is defined via the
relation $\mathbf{\hat{s}}_{i} = 2(\mathbf{s}^{2}_{i}-\mathbf{s}^{1}_{i})/a \equiv
\{\cos{\phi_{i}}, \sin{\phi_{i}} \}$.  The force $\mathbf{F}_{ij}$ and torque $\tau_{ij}$
exerted on void section $j$ by void section $i$ are given by
\begin{SIequation}
  \label{eq:forceAndTorque}
  \mathbf{F}_{ij} =
  F(|\mathbf{s}^{\text{CoM}}_{j}-\mathbf{s}^{\text{CoM}}_{i}|)\shat_{ij}; \qquad \tau_{ij}
  = \frac{1}{4} \sum_{m,n = 1}^{2} ((\mathbf{s}^{n}_{j}-\mathbf{s}^{\text{CoM}}_{j}) \times \mathbf{F}_{ij}^{mn}),
  \end{SIequation}
where $\shat_{ij}
=(\mathbf{s}^{\text{CoM}}_{j}-\mathbf{s}^{\text{CoM}}_{i})/|\mathbf{s}^{\text{CoM}}_{j}-\mathbf{s}^{\text{CoM}}_{i}|$
and $\mathbf{F}_{ij}^{mn}$ is the force from the point at $\mathbf{s}^{m}_{i}$ on the
point at $\mathbf{s}^{n}_{j}$. Physically, this is equivalent to modeling a non-axisymmetric
void as two axisymmetric voids that are rigidly attached to one another in the $x=0$ plane.

To compare in detail the predictions of the above model with observed chromosome/void
configurations in the metaphase plate, we performed simulations to find configurations of
$\nChr$ identical dimers in mechanical equilbrium, given a specific repulsive potential
$U(s)$ and a randomized initial configuration of dimers within with an experimentally
determined spindle boundary.\footnote{A set of dimers may have many such equilibria,
corresponding to different local minima of the total energy, and our simulations do not in
general find the dimer configuration that globally minimizes the total energy of the
system. However, all mechanical equilibria we find generate similar
configuration statistics (S.I.~Fig.~\ref{fig:simComparison}).} To do this, we used
Eqn.~\ref{eq:forceAndTorque} to numerically integrate the over-damped equations of motion
for the dimer center of mass and
orientation,
\begin{equation*}
 \mathbf{\dot{s}}^{\text{CoM}}_{j} = \frac{1}{\gamma} \Bigg( \sum_{i\neq j} \mathbf{F}_{ij} +
\mathbf{F}_{\text{bound}, j} \Bigg); \qquad   \dot{\phi_{j}} = \frac{1}{\gamma_{\phi}} \sum_{i\neq j} \mathbf{\tau}_{ij},
\end{equation*}
discretized as
\begin{SIequation}
   \label{sieq:EoMs}
\mathbf{s}^{\text{CoM}}_{j}(t+\Delta t) = \mathbf{s}^{\text{CoM}}_{j}(t) + \frac{\Delta t}{\gamma} \Bigg( \sum_{i\neq j} \mathbf{F}_{ij}(t) + \mathbf{F}_{\text{bound}, j}(t) \Bigg); \qquad
\phi_{j}(t+\Delta t) = \phi_{j}(t) + \frac{\Delta t}{\gamma_{\phi}} \sum_{i\neq j} \mathbf{\tau}_{ij}(t),
\end{SIequation}

where $\mathbf{F}_{\text{bound}, j}$ is a force exerted by the boundary on dimer $j$, and
we chose the drag coefficients $\gamma$ and $\gamma_{\phi}$ and time step $\Delta t$ such
that for later time-steps $|\Delta \mathbf{s}^{\text{CoM}}_{j}| \ll a/2$ and $\Delta \phi_{j} \ll
\pi$. In early time-steps, randomly positioned particles can be arbitrarily close together and so
can experience very large forces and torques; we therefore explicitly limit the maximum
displacement of each component of a dimer's center of mass to $\Delta x_{\text{max}}$ and
its angular rotation to $\Delta \phi_{\text{max}}$; this only
affects the first few time-steps of the evolution. In Eqns.~\ref{sieq:EoMs}, the term  $\mathbf{F}_{\text{bound}, j}$ is
included to prevent repulsive dimers escaping the metaphase plate. We implement this
confining force by placing $N_{\text{bound}}$ equally-spaced points $\mathbf{b}_{k}$ along
the experimentally determined boundary of the metaphase plate
(S.I.~Fig.~\ref{fig:repulsiveSims}(b)), each of which repels the dimer $j$ with a force
derived from the Yukawa potential $ U^{B}(s) = B_{0}e^{-s/\lambda_{B}}/s$,
\begin{equation*}
\mathbf{F}_{\text{bound}, j} = \sum_{k=1}^{N_{\text{bound}}} \mathbf{F}_{kj}^{B} ;
\qquad \mathbf{F}_{kj}^{B} = - \frac{\partial U^{B}(s_{kj})}{\partial s_{kj}}\shat_{kj},
\end{equation*}
where $s_{kj}$ is the distance between boundary particle $k$ and the center of mass of
dimer $j$, $\shat_{kj}$ is the unit vector along the line joining these points, and the
parameters $B_{0}$ and $\lambda_{B}$ determine the amplitude and range of the Yukawa
repulsion. We choose $B_{0}$ and $\lambda_{B}$ to ensure that the average area of the
convex hull of the final configuration of ellipses matches, within 10\%, the average area of the convex
hull of the experimentally observed chromosome sections.

We time-evolve the initial configurations using Eqn.~\ref{sieq:EoMs} until the absolute
values of the displacements of the particles comprising all dimers sum to a value less
than a tolerance $tol \ll a/2$. The values of all parameters used in our simulations are
given in S.I.~Table~\ref{tab:simParams}. Finally, to convert the output of a simulation
into a form that can be compared with experimental data, we replace the dimers with
ellipses whose shape parameters $a$ and $b$ are derived from the specific experimental
data set that was used in that simulation (S.I.~Fig.~\ref{fig:repulsiveSims}(c)).

\begin{table}[h!]
\begin{tabular}{|c|l|l|l|l|}
\hline
                                                              &  Parameter & Description & Value & Unit \\ \hline \hline
\multirow{4}{*}{\makecell{all \\ sims}}  & $N_{\text{bound}}$ & number  of  boundary  points&  200 &  1 \\ \cline{2-5}
                                                              & $B_{0} \Delta t/ \gamma$ & effective amplitude of boundary repulsion & $10^4$  &  $\um^{3}$ \\ \cline{2-5} 
                                                              &  $\lambda_{B}$ & range of boundary repulsion   & 0.1  & $\um$ \\ \cline{2-5}
                                                              &  $tol$ &
                                                                         \makecell[l]{simulation
                                                                         stops when total
  \\ motion is less than this value} & $10^{-5}$ & $\um$\\ \cline{2-5}
                                                                &  $\Delta x_{\text{max}}$ &
                                                                         \makecell[l]{largest  $\yhat$ or  $\zhat$ displacement each \\ dimer can make in a time-step} & $10^{-2}$ & $\um$\\ \cline{2-5}
   &  $\Delta \phi_{\text{max}}$ &\makecell[l]{largest angular displacement each \\
  dimer can make in a time-step} & $10^{-1}$ & rad \\ \hline \hline
  \multirow{3}{*}{\makecell{$U(s) = U_{\text{int}}(s)$ \\ (Eqn.~\ref{eq:Uint})}} &  $\alpha$ & +1/2 spacing in defect quadrupole & 9.9  & $\um$ \\ \cline{2-5} 
                                                                                                 & $\beta$   & -1/2 spacing  in defect quadrupole  & 11.0  & $\um$ \\ \cline{2-5} 
                                                                                                 &
                                                                                                   $ 8\pi k \Delta t/\gamma$   & effective elastic constant & 1.26 & $\um^{2}$  \\ \hline \hline 
  \multirow{1}{*}{$U(s) = A_{3} s^{-3}$}  & $A_{3} \Delta t/ \gamma$ & effective amplitude of repulsion & $10^{-1}$ &$\um^{5}$ \\ \hline \hline
    \multirow{1}{*}{$U(s) = A_{6} s^{-6}$}& $A_{6} \Delta t/ \gamma$ & effective amplitude of repulsion & 0.5 & $\um^{8}$ \\  \hline \hline
  \multirow{2}{*}{$U(s) = A_{Y} e^{-s/\lambda_{Y}}s^{-1}$}& $A_{Y} \Delta t/ \gamma$ & effective amplitude of repulsion & 0.02 & $\um^{3}$ \\ \cline{2-5} 
                                                                                                &  $\lambda_{Y}$ & range of Yukawa repulsion   & 1  & $\um$ \\ \hline \hline
\end{tabular}
\caption{Parameters for repulsive ellipse simulations. An entry of ``1'' in the Unit
column indicates that a parameter is unitless. In all simulations, we set $\gamma_{\phi} =
(\frac{a}{4})^{2}\gamma$.}
\label{tab:simParams}
\end{table}

\subsection{Results for Specific Long-Range Repulsive Potentials}
The framework provided in S.I.~\ref{sec:repulsiveEllipses} applies for arbitrary repulsive
interactions in the $x=0$ plane. To obtain the results presented in Main Text Fig.~4(e),
we used the potential $U(s) = A_{6} s^{-6}$, where the value of $A_{6}$ and other
simulation paramerers are given in S.I.~Table~\ref{tab:simParams}. We also used the
arbitrarily chosen potentials $A_{3} s^{-3}$, $B_{1} e^{-s/\lambda_{B}} s^{-1}$, and the
potential $U_{\text{int}}(s)$ given in Eqn.~\ref{eq:Uint} derived from the $xy$-plane
model presented in S.I.~\ref{sec:voidInt2D}, with the parameter $s$ now interpreted as the
distance between two points in the $yz$-plane. All simulation parameters are given in
S.I.~Table~\ref{tab:simParams}.

\subsection{Comparison of Experimental and Simulated Configurations}
\begin{figure}[t!]
  \centering
  \includegraphics[width=\textwidth]{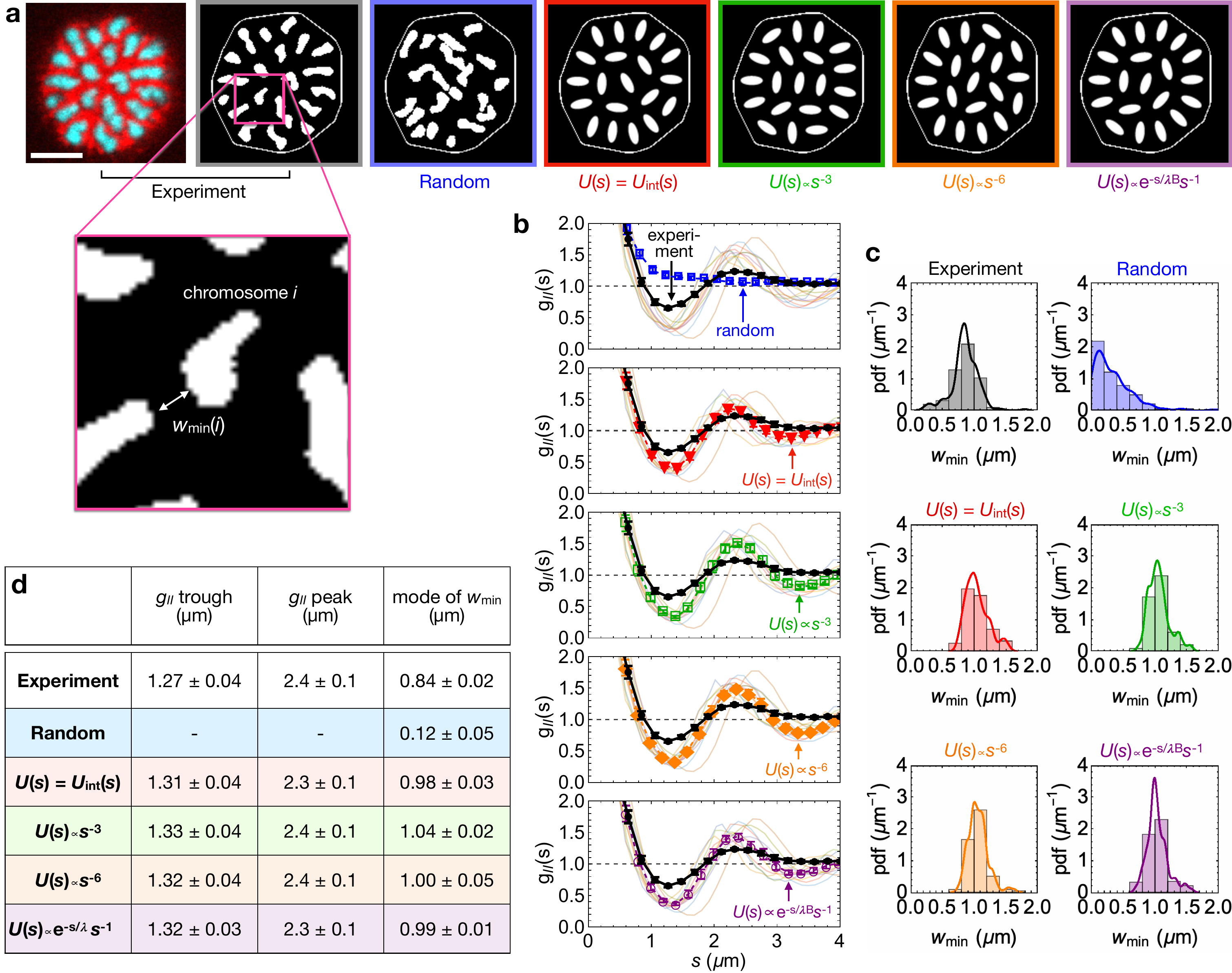}
  \caption{\label{fig:simComparison} Comparison of simulated chromosome configurations
with experimentally observed ones. (a) From left to right: original and binarized
experimental images of chromosomes in the metaphase plate (scale bar $5 \um$); final
configuration of a simulation where we randomized the positions of the binarized
experimental chromosome sections; and the final configurations of four simulations of
ellipses interacting via long-range repulsive potentials. All simulations are performed
using the experimentally measured spindle boundary (white curve). In the long-range
repulsion simulations, the major and minor axes of the ellipses shown are determined by
the average experimentally measured chromosome shapes. \textit{Zoom:} For chromosome $i$,
the nearest neighbor separation $w_{\text{min}}(i)$ is the smallest distance between the
surface of chromosome $i$ and the surface of any other chromosome. (b) Intensity-intensity
correlation function $g_{II}(s)$ for all simulations, plotted alongside $g_{II}(s)$ for
binarized experimentally observed chromosome sections. In each plot, one faint curve
corresponds to one run of a simulation, with boundary geometry and ellipse shape derived
from a specific experimental data set, as shown in (a). Colored points with error bars
represent averages of simulated data; black points with error bars show the average
experimental $g_{II}(s)$ (Main Text Fig.~4(b)). All error bars indicate SE. (c) Histograms
of $w_{\text{min}}$ for all identified chromosomes in the binarized experimental images
($n=216$ chromosomes in 11 spindles), the corresponding randomized chromosome sections,
and the images generated by our repulsive ellipsoid simulations. (d) Table of chromosome
configuration parameters for the binarized experimental images and all simulations. The
randomized chromosome simulations lack local extrema in $g_{II}(s)$, so the first two
parameters are not defined for this data set.}
\end{figure}

It is immediately visually evident that simulated random chromosome configurations are
very different to experimentally observed configurations, while all four repulsive
ellipsoid simulations produce configurations that share several features with experimental
ones, including well-separated chromosomes/ellipses, and the formation of a ring of
$\sim$15 mostly radially oriented chromosomes ellipses near the spindle boundary
(S.I.~Fig.~\ref{fig:simComparison}(a)). To quantitively compare these configurations, we
measured three parameters: the positions of the trough and peak in $g_{II}(s)$ and the
minimum distance $w_{\text{min}}(i)$ between the surface of chromosome section $i$ and the
surface of its nearest neighbor (S.I.~Fig.~\ref{fig:simComparison}(a, \textit{Zoom})). In
terms of these parameters, randomized configurations are very different from
experimentally observed ones (Main Text Fig.~4(c) \&
S.I.~Fig.~\ref{fig:simComparison}(b-d)), while configurations derived from the repulsive
ellipsoid simulations are similar (S.I.~Fig.~\ref{fig:simComparison}(d)). We note,
however, that in the simulated repulsive data, peaks and troughs in $g_{II}(s)$ have a
larger amplitude than in the experimental $g_{II}(s)$, while the modal value of
$w_{\text{min}}$ is $\sim$10\% higher. These discrepancies presumably occur because the
dimers in each simulation are of identical size and shape, whereas chromosomes/voids in a
real spindle is not; this may facilitate more efficient ordering of simulated dimers
compared to metaphase chromosomes~\cite{Li2016b}.
  
\newpage

\section{Analytical Model for Repulsion Between Voids in 2D Nematic ($z=0$ Plane Model)}
\subsection{Expression for Deformation Field Around Parallel Voids \& Associated
  Interaction Energy}
\label{sec:voidInt2D}
In a 2D nematic with equal elastic constants, the general formula for the
orientation of the director in the vicinity of a collection of topological defects of
strength $\sigma_{i}$ positioned at $z_{i} \equiv x_{i}+i y_{i}$ is given by
the formula~\cite{Vafa2020}:
\begin{equation*}
  \theta(z, \bar{z}) = i \sum_{i}\sigma_{i}\log{\frac{\bar{z} -\bar{z_{i}}}{z-z_{i}}} + 2\psi,
  \end{equation*}
where the constant $\psi$ controls the director orientation at infinity; setting $\psi=0$
realizes our desired far-field value $\theta=0$. The same reference gives the
corresponding energy, under the assumption that the defects are separated by distances
much larger than the core size $a_{0}$, 
\begin{equation*}
U = -8\pi k \sum_{i \neq j}\sigma_{i}\sigma_{j} \log{\frac{z-z_{i}}{a_{0}}} + C,
\end{equation*}
where the constant $C$ includes the core energies of all defects. 

We describe the deformation field due to an isolated 2D void as that generated by a single row of
four defects, comprising two $-1/2$ defects separated by a distance $\beta$ and two $+1/2$
defects separated by a distance $\alpha$, with $\alpha < \beta$. All defects lie along the
line $y=0$ (Main Text Fig.~4(d, \textit{left}) \&
S.I.~Fig.~\ref{fig:voidDeformationModel}). The explicit expression for the director
orientation (thin gray curves in figures) is
\begin{equation*}
  \begin{split}
  \theta(x,y) = \frac{1}{2} \Bigg{(}
  -\arg\Big{(} 1 + \frac{4 i y}{-2 x - 2 y +\alpha}\Big{)}
  -\arg\Big{(} 1 + \frac{4 i y}{2 x + 2 y +\alpha}\Big{)} \\
  +\arg\Big{(} 1 + \frac{4 i y}{-2 x - 2 y +\beta}\Big{)}
  +\arg\Big{(} 1 + \frac{4 i y}{2 x + 2 y +\alpha}\Big{)}  \Bigg{)}
  \end{split}
\end{equation*}
and the corresponding energy is given by
\begin{equation*}
  U_{1} = -4\pi k\log{ \frac{16 a_{0}^{2} \alpha \beta}{(\alpha^{2}-\beta^{2})^{2}}}+C.
  \end{equation*}
For two parallel rows of defects, both centered at $x=0$ and separated by
a vertical distance $d$ (Main Text Fig.~4(d, \textit{left})), the total energy is given by
\begin{equation*}
  U_{2}(d) =  -8\pi k \Bigg( 2 \log{\Big(\frac{d}{a_{0}}\Big)} + \log{\Big( \frac{256 a_{0}^{4}\alpha \beta
      \sqrt{(\alpha^{2} + d^{2})(\beta^{2} + d^{2})}}{(\alpha-\beta)^{2}(\alpha+\beta)^{2}(\alpha^{4} -2\alpha^{2}(\beta^{2}-4
      d^{2}) + (\beta^{2}+4 d^{2})^{2})} \Big)} \Bigg) +2\,C,
\end{equation*}
and the energy due to interactions between the voids is given by
\begin{SIequation}
  \label{eq:Uint}
  U_{\text{int}}(d) = U_{2}(d)-2U_{1} =  -8\pi k\log{\Big( \frac{16 d^{2} \sqrt{(\alpha^{2} + d^{2})(\beta^{2} +d^{2})}}{(\alpha^{4} -2\alpha^{2}(\beta^{2}-4 d^{2}) + (\beta^{2}+4 d^{2})^{2}}\Big)}. 
\end{SIequation}

\subsection{Interaction Between Voids is Always Repulsive}
Evaluating the derivative $\text{d}U_{\text{int}}/\text{d}d$ and completing all available squares
shows that $\text{d}U_{\text{int}}/\text{d}d <0 $ for all positive values of $\alpha, \beta,$ and $d$.

\subsection{Relationship Between Void Boundary Shape and Spacing Between Topological Defects}
\begin{figure}[t!]
  \centering
  \includegraphics[width=0.8\textwidth]{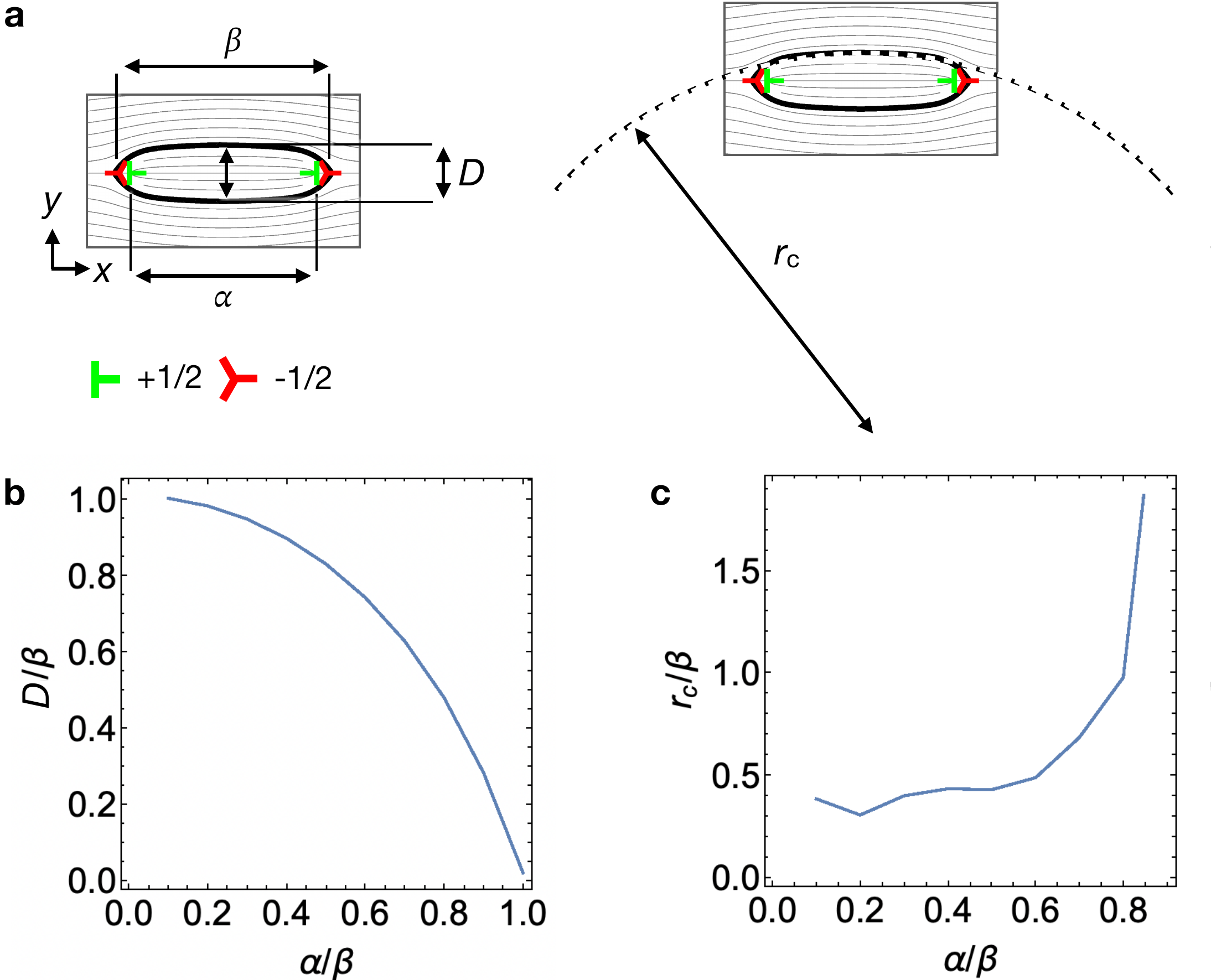}
  \caption{\label{fig:voidDeformationModel} Relationship between void boundary geometry
    and defect spacing. (a) For a void in our 2D model, the defect spacing parameters
    $\alpha$ and $\beta$ uniquely determine the void waist width $D$ and radius of
    curvature $r_{c}$. (b \& c) Plots of void aspect ratio $D/\beta$ and  waist
    curvature $r_{c}/\beta$ as a function of inner defect spacing $\alpha$.}
\end{figure}

Given specific values of $\alpha$ and $\beta$, we define voids in our model as the 2D
regions enclosed by those integral curves of the vector field $\{
\cos{\theta(x,y)},\sin{\theta(x,y)} \}$ that pass through both -1/2 defects (Main Text
Fig.~4(d, \textit{top left}), thick black curve). This construction yields tactoid-like
void shapes while ensuring that tangential boundary conditions are satisfied on the void
surfaces. However, since the integral curves are not exact circle arcs, we must choose
the void spacing parameters $\alpha$ and $\beta$ to match the experimentally observed
tactoid shape (Main Text Fig.~3(f)). To determine the values of the parameters of $ U_{\text{int}}(s)$
given in Table~\label{red:simParams}, we set $\beta$ equal to the
experimentally measured void length ($\beta = 11.0 \um$), and choose the parameter
$\alpha$ such that the curvature of the deformation field at the void boundary matches the
radius of curvature of the experimentally observed voids ($\alpha = 9.9 \um$;
S.I.~Fig.~\ref{fig:voidDeformationModel}).

\newpage

\bibliographystyle{unsrt}
\bibliography{/Users/colmkelleher/Documents/latex_harvard/orientation_paper_draft/library.bib}\clearpage

 \clearpage

 \clearpage

 \clearpage

 \clearpage

 \clearpage

 \clearpage

 \clearpage

 \clearpage

 \clearpage

 \clearpage

 \clearpage

\clearpage

\clearpage

